\documentclass[final,5p,times,twocolumn]{elsarticle}
\usepackage{hyperref}
\usepackage{amssymb}
\usepackage{subfigure}
\usepackage{epsfig}
\usepackage{epstopdf}
\usepackage{algorithm}
\usepackage{xcolor}
\usepackage{graphicx}
\usepackage{amsmath}
\usepackage{algorithm}
\usepackage[noend]{algpseudocode}

\usepackage{lineno}

\biboptions{authoryear}

\makeatletter
\def\BState{\State\hskip-\ALG@thistlm}
\makeatother

\begin{document}

\begin{frontmatter}



\title{Impact of the purposeful kinesis on running waves}


\author{N. \c{C}abuko\v{g}lu }
\ead{nc243@le.ac.uk}

\address{Department of Mathematics, University of Leicester, Leicester, LE1 7RH, UK}

\begin{abstract}

The basic model of purposeful kinesis was developed recently (Gorban \& \c{C}abuko\v{g}lu, Ecol. Complex. 2018, 33, 75–83) on the basis of the ``Let well enough alone" idea. According to this model the diffusion drops while the reproduction coefficient is increasing. That is, species prefer to stay in a good condition and the population gives birth; otherwise, in the bad situation individuals want to run away because of the fatal conditions. In this study, we analyse the impact of the purposeful kinesis model on running waves. The running waves in the population with kinesis are studied using numerical experiments. Both monotonic and non-monotonic (Allee effect) dependence of the reproduction coefficient on the population density are studied. The possible benefits of the purposeful kinesis are demonstrated: with the higher diffusion, while the population without kinesis ends up with extinction, the population with kinesis stays alive and has the running wave behaviour. While the kinesis of the prey population is decreasing, the wave amplitude gets smaller. On the other hand, for the lower kinesis of predators they have a sharp increase.

\end{abstract}


\end{frontmatter}

\section{Introduction}

In this study, we have analysed the diffusion model with kinesis. Kinesis is the non-directional movement according to the change of the local environment. That is, the organism gets the information from the living area, and therefore prefer to stay or move to the other beneficial place. Taxis is also related to mobility. However, taxis requires the non-local information. The organism with taxis has the directed movement towards to the condition or opposite side of the stimulus.

We aim to explore the impact of purposeful kinesis on running waves using PDE models. The classical PDE model of population dispersal was proposed by \cite{Patlak1953} and  \cite{Keller1971}. It was used to model taxis behaviour \citep{Hillen2009}. The reaction-diffusion model can be used  for kinesis modelling. The diffusion coefficient can depend on the local situation. \cite{Cosner2014} studied the reaction-diffusion models on animal dispersal. He presented the study on animal movement: which is the better condition to stay or leave, being slow or random movement, increasing the population size is harmful or not.

In this study, we employ the purposeful kinesis model \citep{GorCabuk2018}. Often  `Purposeful' means intentionality that individuals are unable. \cite{Rosenblueth1950} developed the general concept of purposeful behaviour. `Purpose' brings the optimization idea and this concept requires the evolutionary optimality \citep{Parker1990}.

There is a connection between the reproduction rate and diffusion coefficient. Average reproduction coefficient was defined as Darwinian fitness \citep{ Haldane1932, Metz1992, Gorban2007}. Migration should increase Darwinian fitness.

In this paper, we analysed impact of the purposeful kinesis on running waves. In particular, it was presented that the population with Allee effect can spread late but with a higher diffusion the population will stay alive, and in time will continue the running wave behaviour. We did numerical experiments that kinesis model holds spreading invaded area, while the population without kinesis dies. 

\section{Main Results}

In this section, we will give the main results.

\subsection{Running Waves}

The kinesis model was selected in the following form \citep{GorCabuk2018}:
\begin{equation}\label{KinesisModel}
\boxed{
 \partial_t u_ i ( x,t)
   =  D_{0i} \nabla \cdot \left(e ^{-\alpha_i r_i(u_1,\ldots,u_k,s)} \nabla u_i \right)+r_i (u_1,\ldots,u_k,s) u_i,
 }
   \end{equation}
   where:
\begin{itemize}
\item[] $u_i$ is the $i$th species-population density,
\item[]   $s$ is the abiotic characteristics of the living conditions,
\item[] $r_i$ is the reproduction coefficient,
\item[] $D_{0i}>0$ is the equilibrium diffusion coefficient which is defined when the reproduction coefficient is 0,
\item[] $\alpha_i>0$  defines the relation between the diffusion coefficient on the reproduction coefficient.
\end{itemize}

 $D_i=D_{0i}e ^{-\alpha r_i} $ characterizes the diffusion depending on reproduction coefficient. According to that model, the diffusion depends on well-being which is measured by reproduction coefficient.

\cite{GorbanCabukoglu2018} later studied the cost of mobility. The changes of reproduction coefficient give the mobility cost value. If we take into account the cost of mobility then the bifurcation is predicted: when the conditions are getting worse, the mobility increases to a threshold level, and then it vanishes. Therefore, in the worse conditions, there is no solution for mobility. This mobility cost equation can be solved with Lambert $W$-function.

We will present below how the dependence of diffusion on well-being effects the running waves of the population density on space.

Let us compare two models:
\begin{itemize}
\item{ The PDE model for population with the constant diffusion coefficient: without kinesis, KPP (Kolmogorov, Petrovsky and Piskunov, 1937):
\begin{equation}
\partial_t u( t,x) = D \nabla^2 u  + ( 1- u( t,x ) ) u( t,x), \\
\label{Model2}
 \end{equation}
 }
\item{  The PDE model for population  with kinesis (KPP with modified diffusion coefficient \citep{KPP1937}):
\begin{equation}
\partial_t u ( t,x) = D \nabla \cdot \left(e^{-\alpha ( 1- u( t,x ) ) } \nabla u \right) + ( 1- u( t,x ) ) u( t,x), \\
\label{Model1}
 \end{equation}
 }
\end{itemize}

We have first used MATLAB \cite{pdepe2017} function to solve one dimensional system of PDE. Then, we used the MATHEMATICA \cite{NDSolve2014} solver which is applied the Hermite method and Newton's divided difference formula to solve two dimensional system.

We selected the space on the interval $[-50,50]$ with Dirichlet boundary conditions and with the initial conditions:
$$u( 0,x) = \dfrac{1}{1+e^{\lambda x}}.$$ The values of the constants are: $D=1$, $\alpha=1$, $\lambda=10$.

\begin{figure}[!ht]
\centering
\subfigure[]{\includegraphics[width=0.42\textwidth]{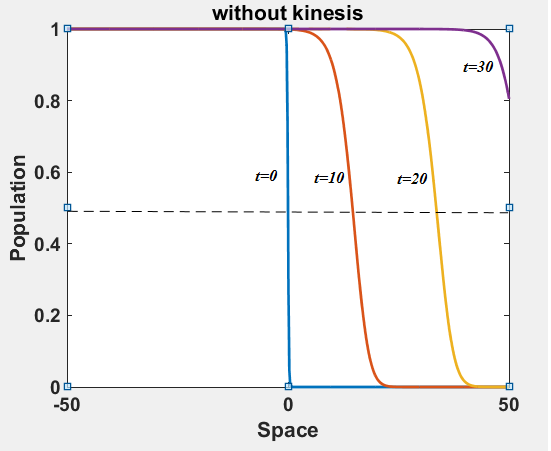}}
\subfigure[]{\includegraphics[width=0.42\textwidth]{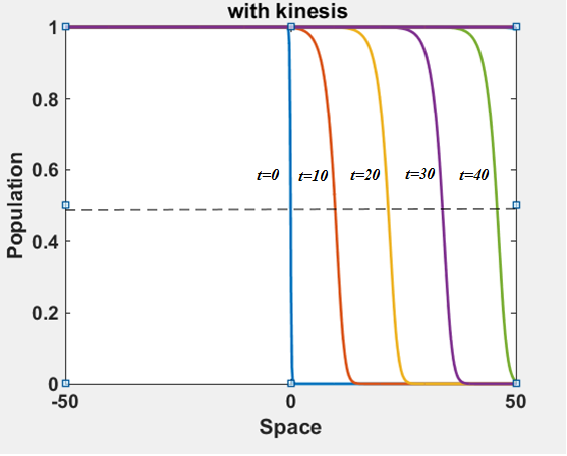}}
\caption{Travelling waves of Population density: a) for animals without kinesis for model (\ref{Model2}) and b) for animals with kinesis for model (\ref{Model1}). The values of the constants are: $D=1$, $\alpha=1$, $\lambda=10$. \label{spread12}}
\end{figure}

In the large time the running waves converge to a unique front with the velocity $v$. For example, the logistic growth model $r(u)=\alpha u \left( 1- u/K \right)$ speed will be $v_0 = 2\sqrt{\alpha D}$.

\begin{figure}[!ht]
\centering\includegraphics[width=0.5\textwidth]{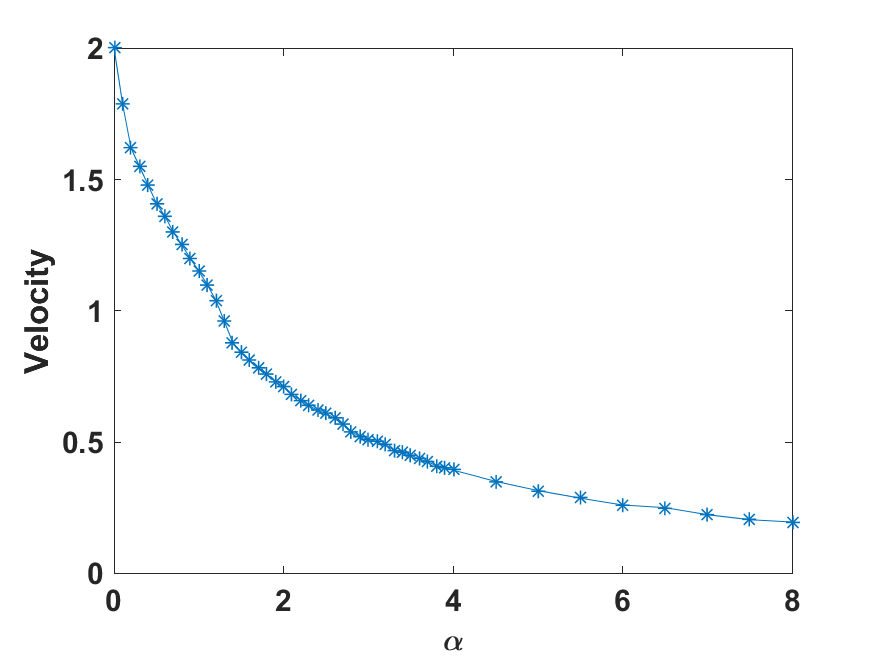}
\caption{Wave velocity as function of the kinesis parameter $\alpha$ for model (\ref{Model1}).\label{velocityalpha}}
\end{figure}

Fig.~\ref{spread12} presents that the proposed `minimal purposeful kinesis model' has the running wave behaviour on space. In these given conditions, the waves in the population model with kinesis is slower which is expalined in Fig.~\ref{velocityalpha}. When we increase the kinesis parameter $\alpha$, the spreading velocity will decrease monotonically.

We have done numerical experiments on MATLAB to see the relations between the wave velocity and the kinesis parameter $\alpha$ (see Fig.~\ref{velocityalpha}). While $\alpha$ is increasing, the velocity starts to decrease monotonically. After $\alpha = 6$, the velocity is almost stabilizing. That is, the population spreading on space can be almost the same in the different time lines. 

\subsection{Running Waves with the Allee effect}

Allee effect was introduced as nonmonotonic behaviour of the reproduction coefficient as a function of population density. At the beginning, the reproduction rate is minimum and  then reaches to the peak. After the maximum reproduction rate, it starts to decrease monotonically while the population size is increasing \citep{Allee1949}. There are a lot of reasons why populations show Allee effect such as low rate feeding \citep{WayBanks1967, WayCammell1970}, and for some reasons the population reduces the predator defence \citep{Kenward1978, Kruuk1964}, give up the breeding for psychological reasons \citep{RalHarLyl1986, Soule1986}, and a wide variety of reasons \citep{CaracoPulliam1984, Folt1987, FosterTreherne1981, TurKar1989}.

\cite{LewisKareiva1993} explored the population spread of asymptotic rates with the Allee effect. In two dimensional analysis, they investigated that the patterns of spread have been influenced by this population model. We have used the coefficients in that study to analyse how the population with kinesis affects the two-dimensional patterns of spread.

The basic form of the reproduction coefficient with Allee effect is $$r(u)=r_0(K-u)( u-\beta).$$ The models with purposeful kinesis on Allee effect  were studied by \cite{GorCabuk2018} previously.

\cite{Gorban1989} introduced the study that when the average population size is less than the optimum density, the evolutionary optimal strategy for populations with the Allee effect collapses with optimal density. There are some other consequences of Allee effect with diffusion: the population spreads in a massive way through the formation, the interaction between species and the movement of separate patches even if the population is inhomogeneous external condition \citep{Morozov2006, Petrovskii2002}.

We may present the reaction-diffusion equations for a single population with  the Allee effect for a system without kinesis (\ref{AlleyWithout}) and  for a system with kinesis (\ref{AlleyWith}).

\begin{equation}
\partial_t u ( t,x) = D \nabla \cdot ( \nabla u ) + k ( 1-u)( u-\beta) u( t,x), \\
\label{AlleyWithout}
\end{equation}
\begin{equation}
\partial_t u( t,x) = D \nabla \cdot \left(e^{-\alpha k ( 1-u) ( u-\beta ) } \nabla u \right)+ k ( 1-u )( u-\beta) u( t,x).\\
\label{AlleyWith}
\end{equation}

The values of constants are: $D=1$, $\alpha=1$, $\beta=0.7$.

Eqs. (\ref{AlleyWithout}) and (\ref{AlleyWith}) are solved for one space variable $x\in[-50,50]$  with  Dirichlet boundary conditions and with the initial conditions:

\begin{equation}\label{1dInitial}
u( 0,x) =\dfrac{1}{1+e^{\lambda x}} ; \lambda=10.
\end{equation}

We choose $k=k(\beta)$ as the normalization constant which was given by \cite{LewisKareiva1993} as determined by the maximum growth rate. $k$ has been determined as 

\begin{equation}
k=1;\label{k1}
\end{equation}
\begin{equation}
k=4/(1-\beta)^2 ;\label{k2}
\end{equation}
\begin{equation}
k=27/\left( 2 \left( \left( \left( 1+\beta \right)^2-9 \beta /2 \right) \left( 1+ \beta \right) + \left( \left( 1+\beta \right) ^2 -3 \beta\right) ^{3/2} \right) \right). \label{k3}
\end{equation}

It is considered that travelling wave solutions to Eqs. (\ref{AlleyWithout}) and (\ref{AlleyWith}) are in the form of $u=U(z)$ with $z=x-vt$, where $v$ is the velocity.

It was shown that there is a unique solution to these Eqs. (\ref{AlleyWithout}) and (\ref{AlleyWith}) with a unique $v$ velocity with the condition $0 < \beta <1$ by \cite{Fife1979}. When the Allee effect parameter is negative, there is a minimum value of velocity such that there exists a corresponding running wave solution  \citep{AronsonWeinberg1978, Fife1979, Hadeler1975}. After a long time period $t \rightarrow \infty$, the solution of the system \ref{AlleyWithout} may converge to a travelling wave solution.

\begin{figure}
\centering
\subfigure[]{\includegraphics[width=0.42\textwidth]{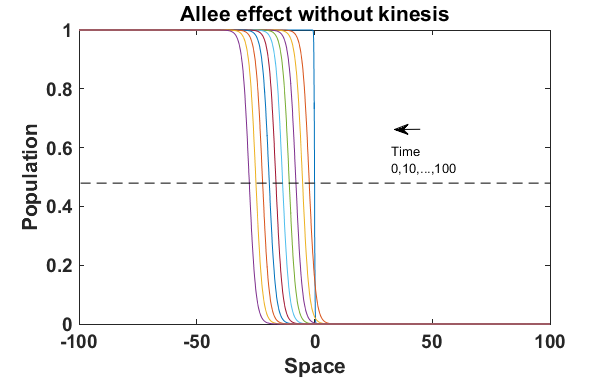}}
\subfigure[]{\includegraphics[width=0.42\textwidth]{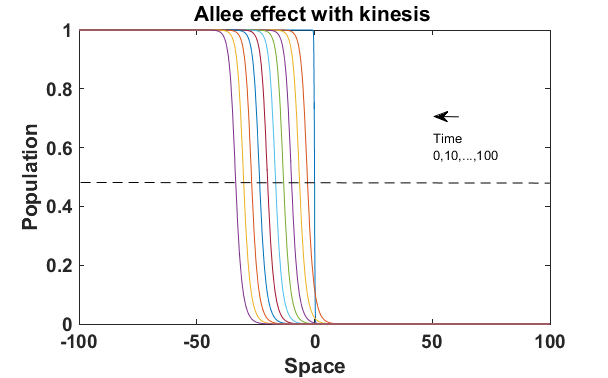}}
\caption{The running waves of population densities with the  Allee effect: (a) for animals without kinesis (\ref{AlleyWithout}), (b) for animals with kinesis in time 0 to 100 (\ref{AlleyWith}). The values of constants are: $D=1$, $\alpha=1$, $\beta=0.7$.  \label{AlleeSpreading1D}}
\end{figure}

The numerical results (Fig.~\ref{AlleeSpreading1D}) show that the population with Allee effect has the running wave behaviour both with kinesis and without kinesis. Population with kinesis demonstrates faster running waves on space. Fast-spreading effect of the population could also end up to extinction for the population with Allee effect. When the Allee effect is higher than the population density, the reproduction coefficient becomes negative. Therefore, this leads the population to extinction.

\begin{figure}[!ht]
\centering
a)\includegraphics[width=0.42\textwidth]{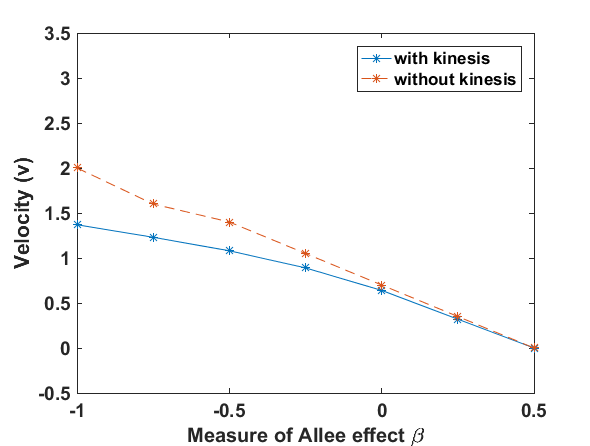}
b)\includegraphics[width=0.42\textwidth]{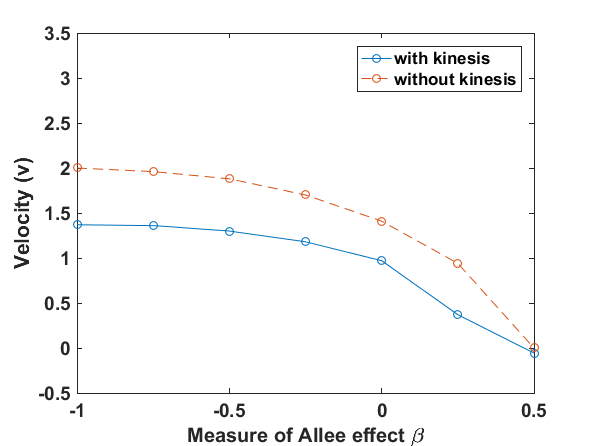}
c)\includegraphics[width=0.42\textwidth]{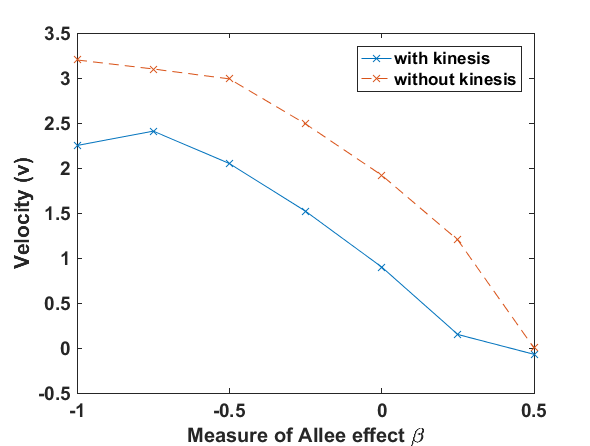}
\caption{ Allee effect parameter and speed $\alpha =1$ and $D=1$. The population with kinesis and without kinesis a) the normalization constant is $k=1$ for models (\ref{AlleyWithout}) and (\ref{AlleyWithout}), b) $k$ is given for model (\ref{k2}). c) $k$ is given for model (\ref{k3}). The system has been solved with the numerical method for the running wave front $u=0.5$ }\label{speedbeta}
\end{figure}

Wave velocity for different models has been displayed as the Fig.~\ref{speedbeta} with the Allee effect parameter $-1 \leq \beta \leq 0.5$. The Allee effect decreases the rate of spread of an invading population. We can see from the Fig.~\ref{speedbeta} that with these conditions the spreading of the population with the Allee effect is decreasing when $\beta$ is getting higher value. Moreover, the spreading velocity of the population without kinesis is faster than the population with kinesis.

\subsection{Two-Dimensional Spread}

Analytical representation of two-dimensional system may be too complex to solve. \cite{AronsonWeinberg1978} presented that the planar travelling waves exist  and this form can be used by an invading population. They studied the two-dimensional population model with Allee dynamics to see the invaded area of the population. The population, with Allee effect, may end up with extinction. This can happen even if the population density exceeds the threshold. The reason is that reproduction, the population growth, cannot be sufficient to prevent the population dispersal. There are some critical factors to determine the invading success: shape and size of the environment. It has been already shown that how the population stay alive between invaded areas and unoccupied places \citep{LewisKareiva1993}.

Now, in this section, we present the two-dimensional model that we want to see the invaded and unoccupied areas of the population with kinesis and without kinesis. In addition to that model, we discuss the population with Allee effect.

\begin{figure}[ht!]
a)\includegraphics[width=0.42\textwidth]{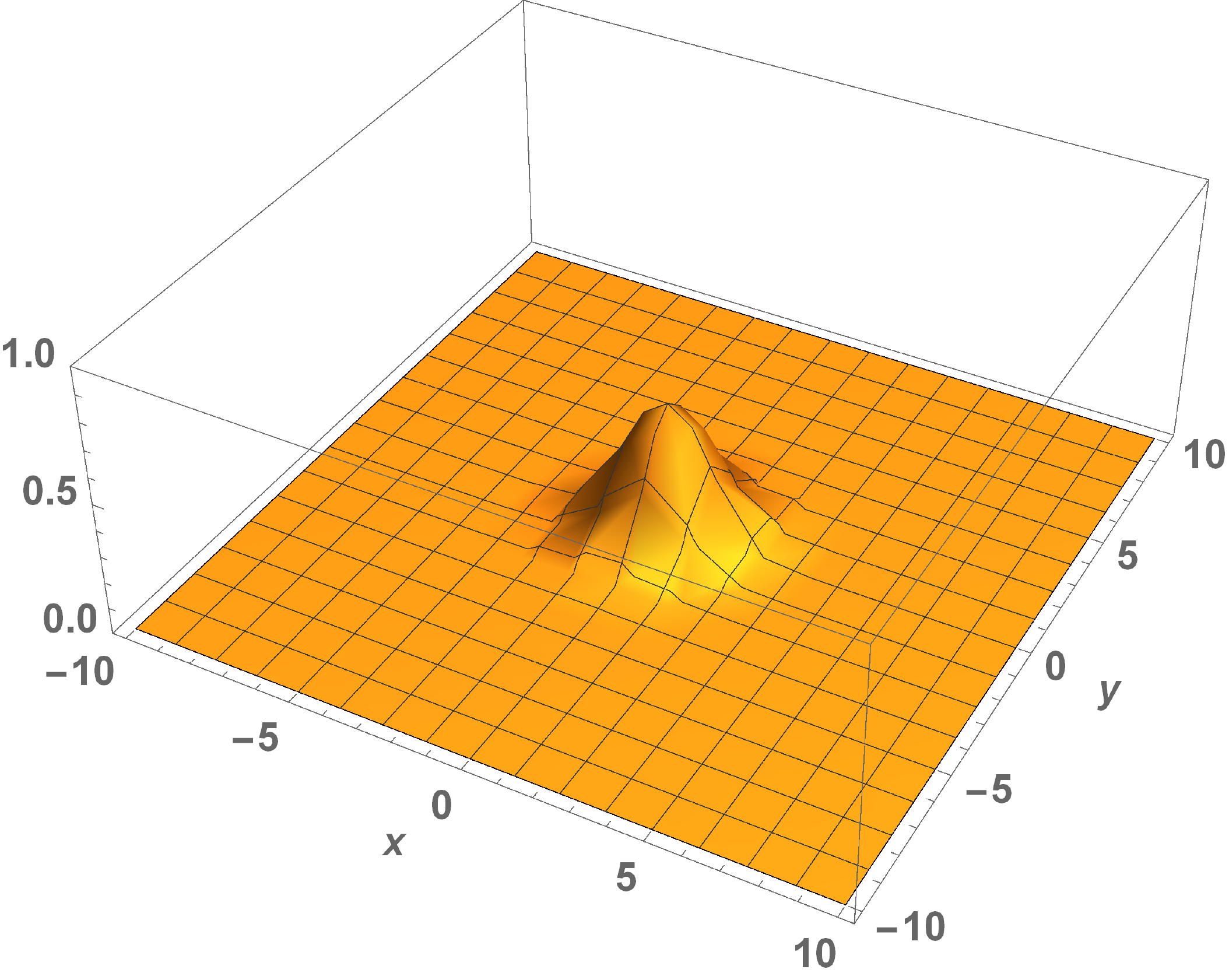}
b)\includegraphics[width=0.42\textwidth]{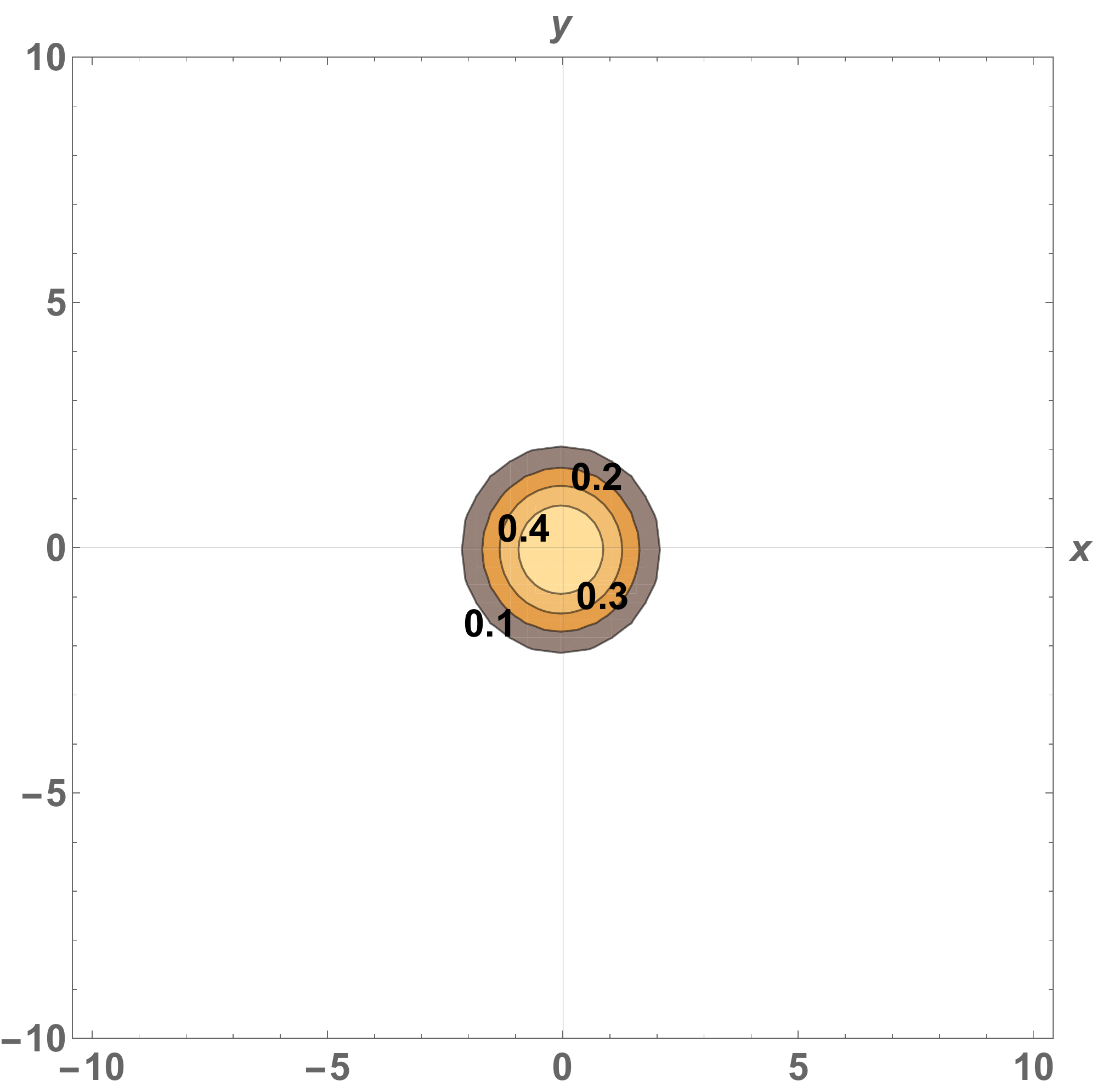}
\caption{Initial distribution ($t=0$): $u( 0,x) =\frac{1}{1+ e^{\frac{x^2+y^2}{2}}}.$ The invaded area by the population with kinesis and without kinesis.\label{initial}}
\end{figure}

Fig.~\ref{initial} shows the initial distribution that the invaded area is the same for the population with kinesis and without kinesis.

\begin{figure}
\centering
a)\includegraphics[width=0.42\textwidth]{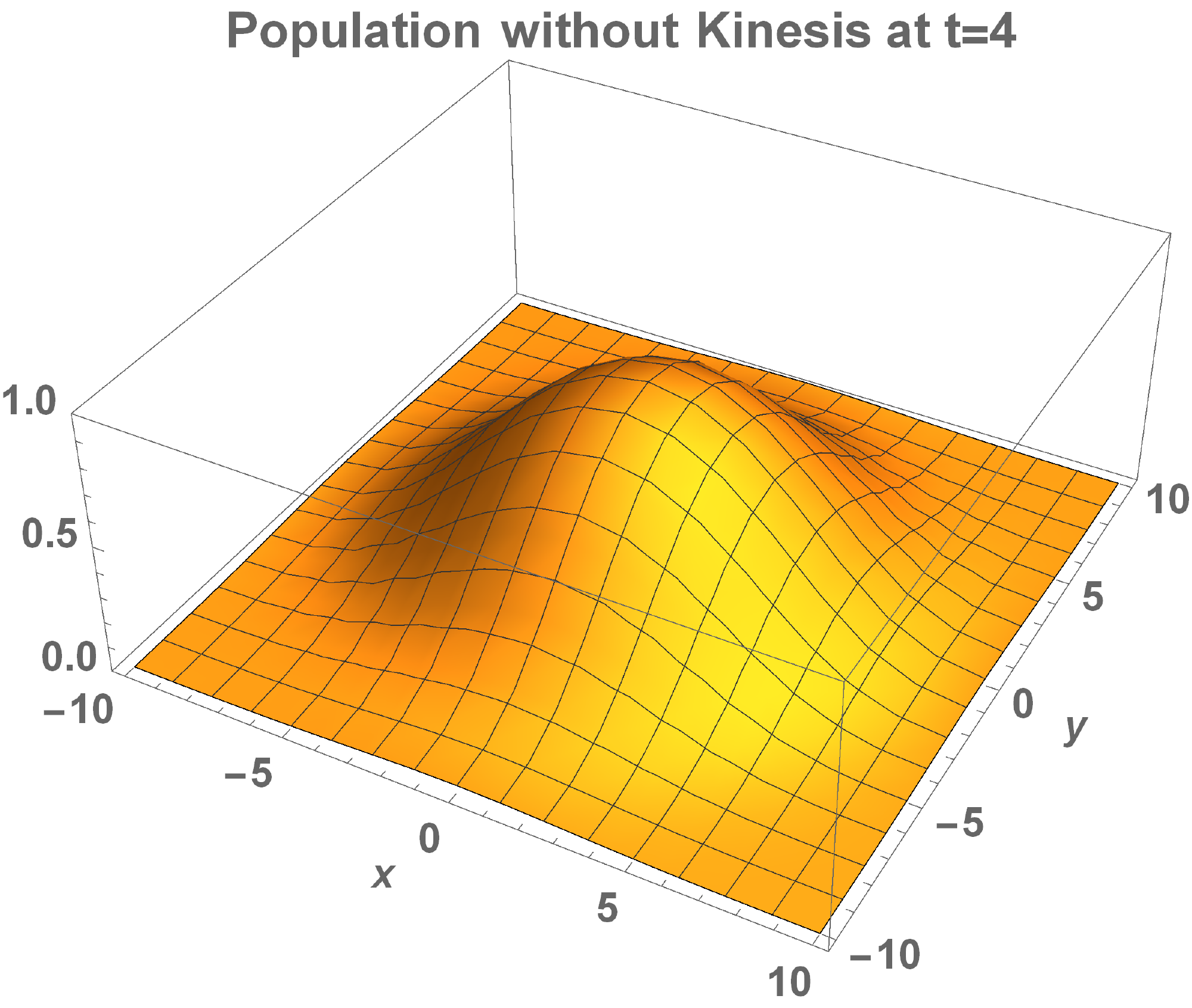}
b)\includegraphics[width=0.42\textwidth]{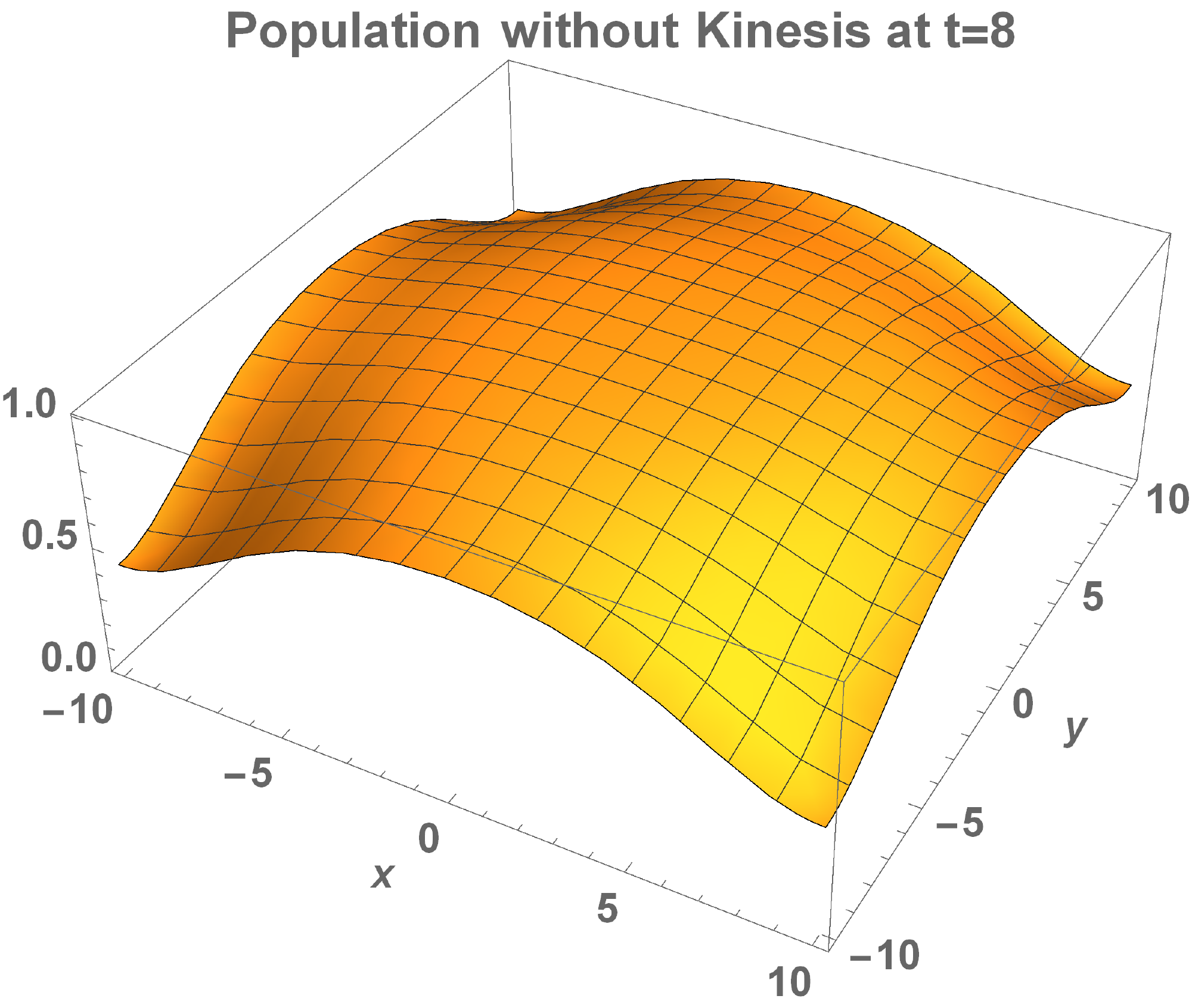}
c)\includegraphics[width=0.42\textwidth]{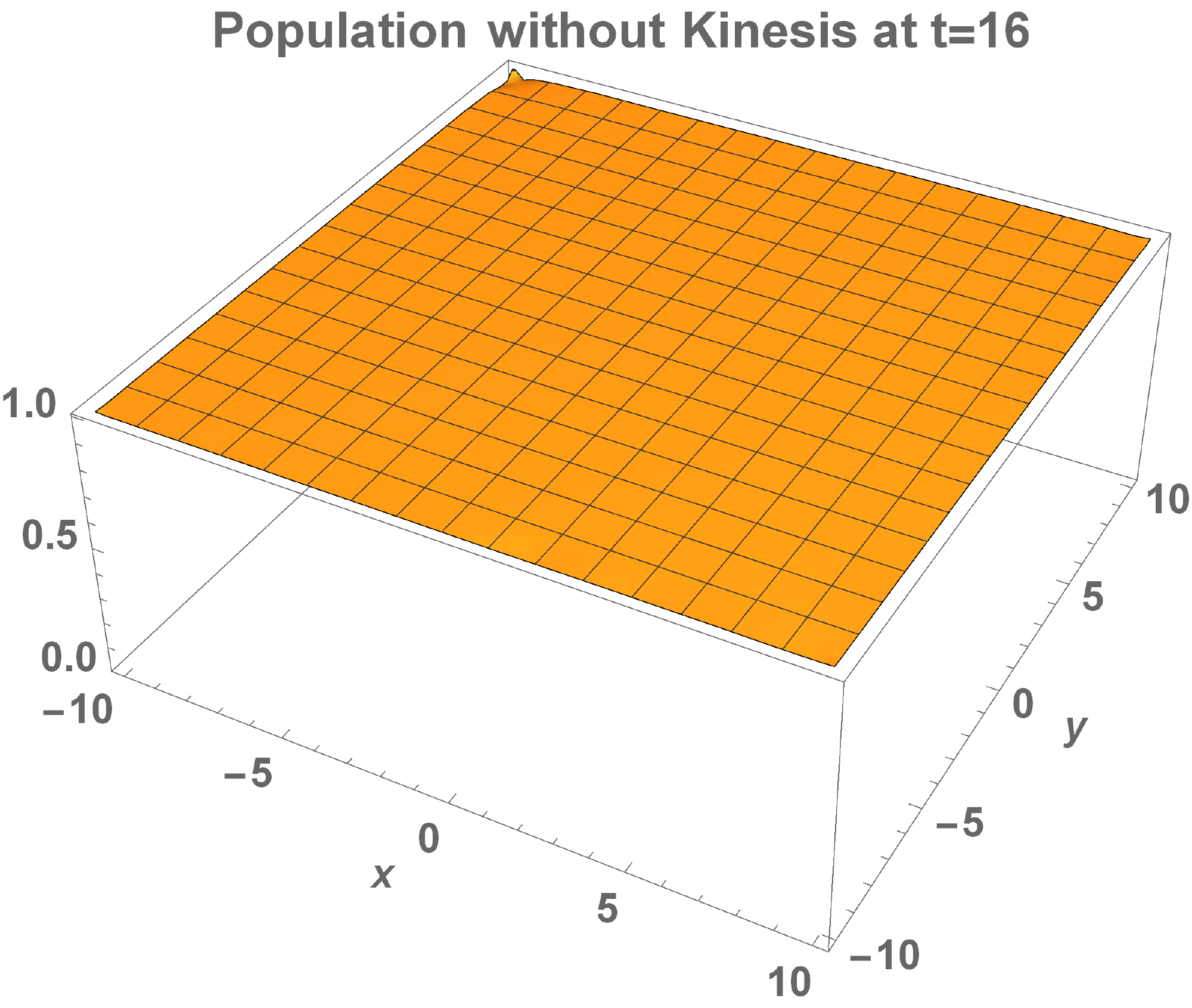}
\caption{ 2D Population {\em without kinesis}  $\alpha =1$ and $D=1$.\label{popwithoutkinesis}}
\end{figure}

\begin{figure}
\centering
a)\includegraphics[width=0.42\textwidth]{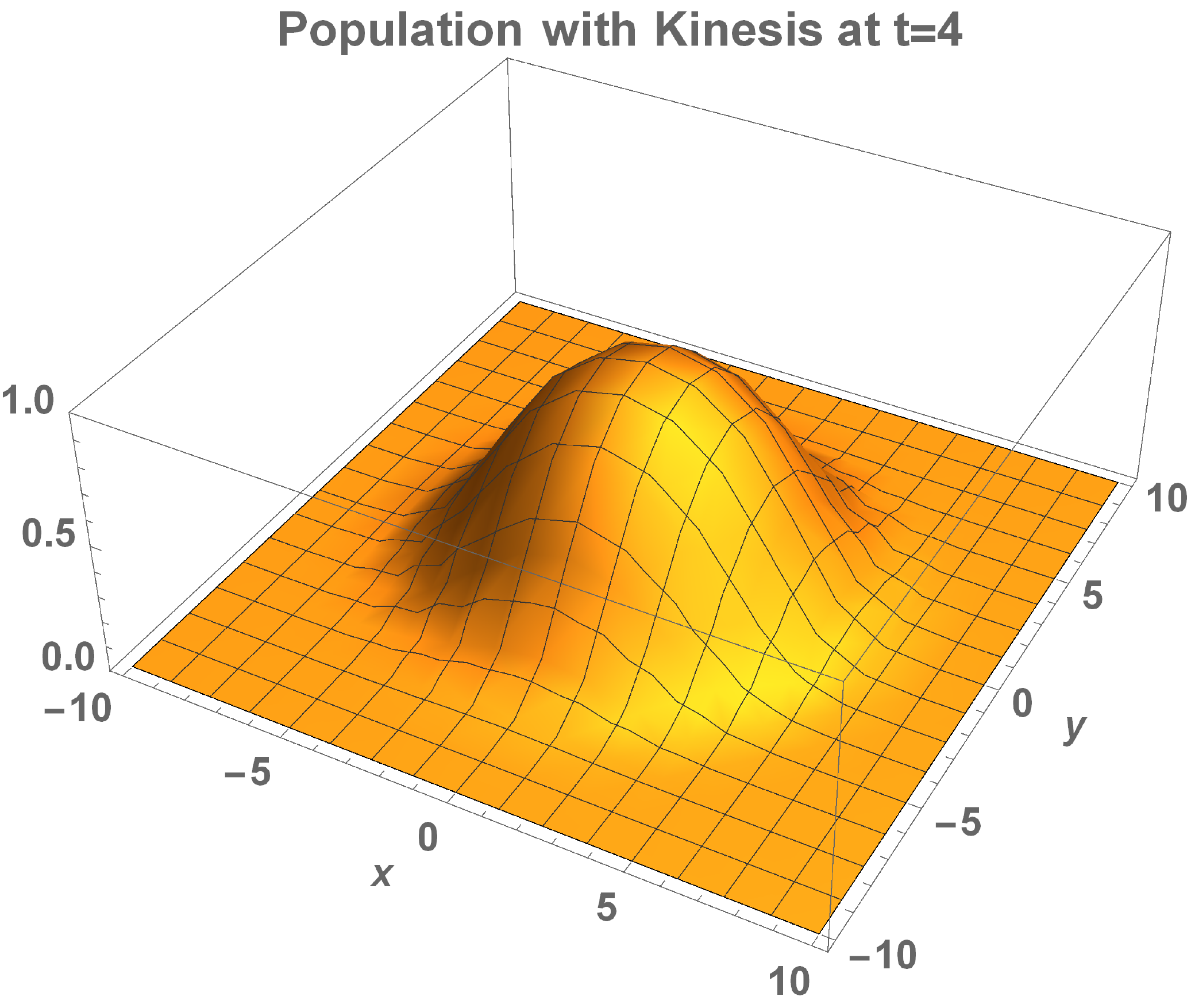}
b)\includegraphics[width=0.42\textwidth]{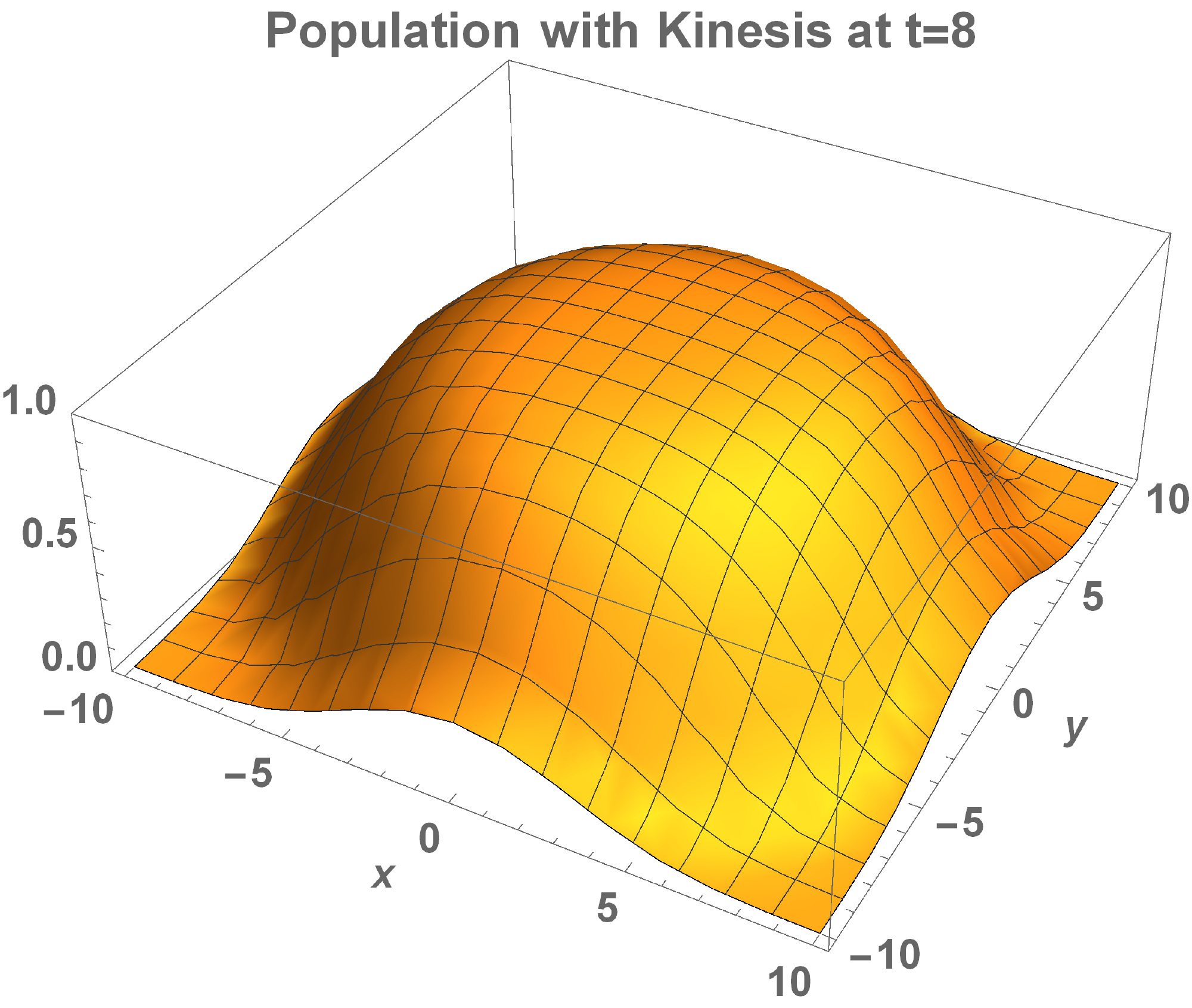}
c)\includegraphics[width=0.42\textwidth]{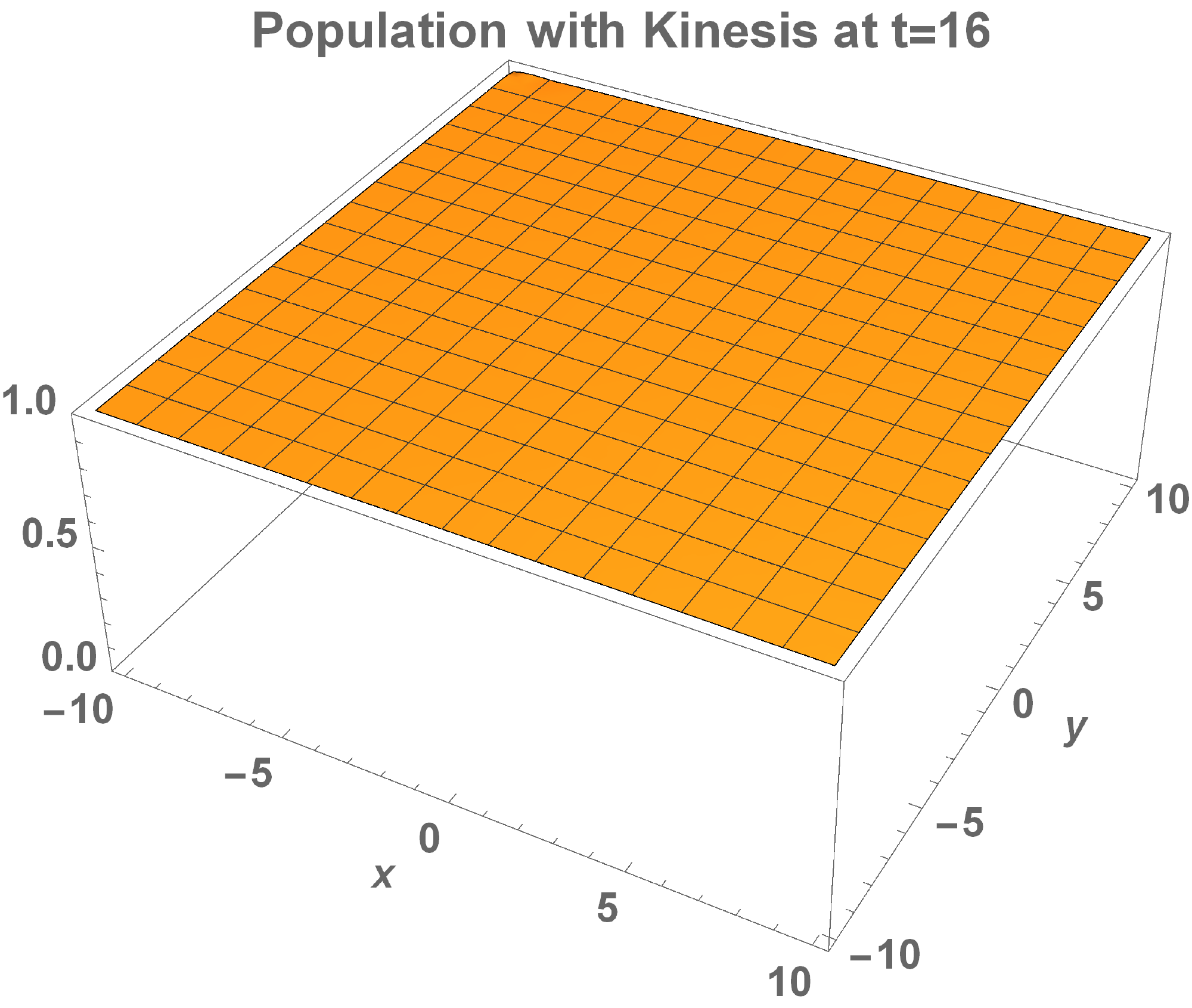}
\caption{ 2D Population {\em with kinesis} $\alpha =1$ and $D=1$.\label{popwithkinesis}}
\end{figure}

\begin{figure}
\centering
a)\includegraphics[width=0.42\textwidth]{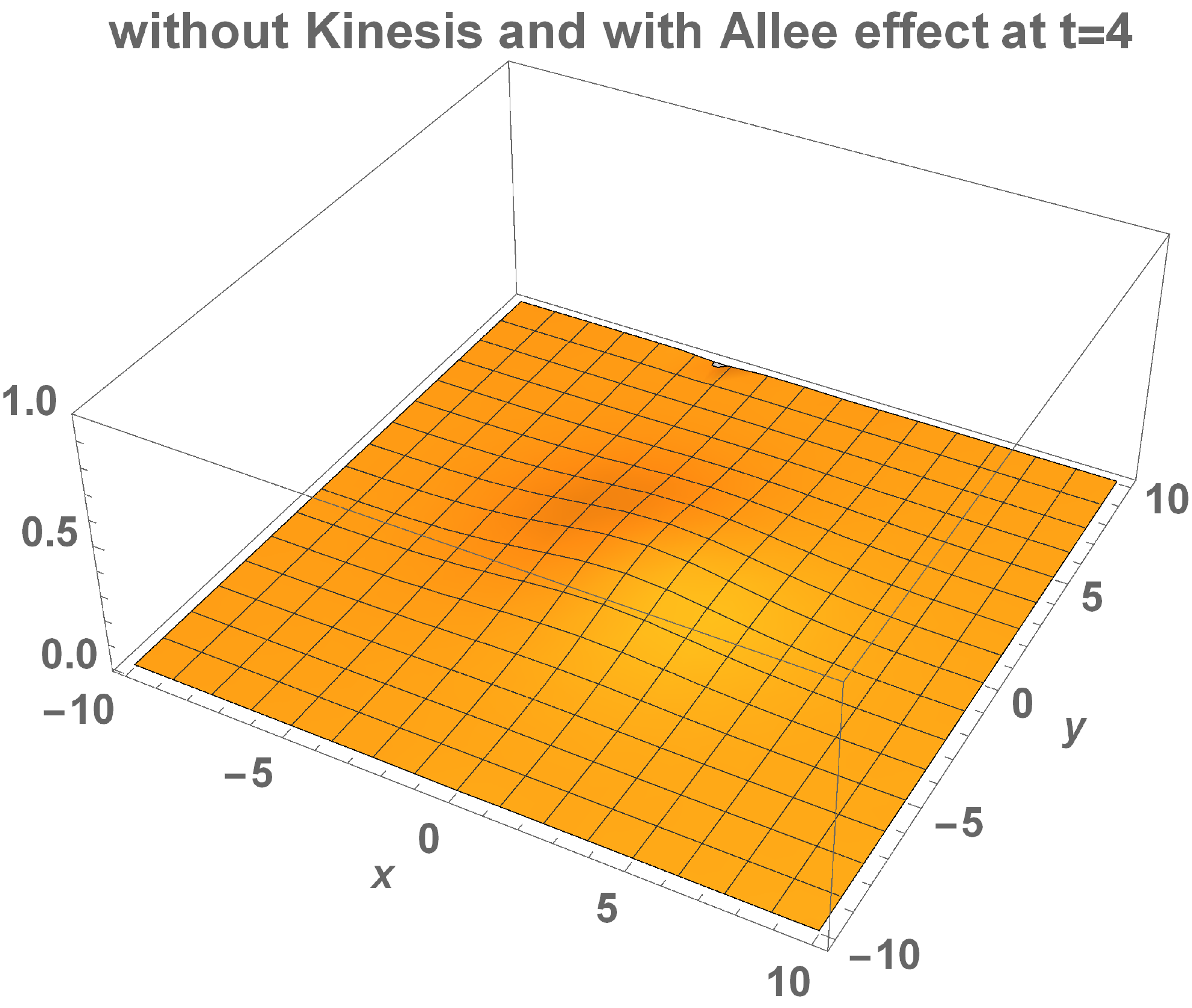}
b)\includegraphics[width=0.42\textwidth]{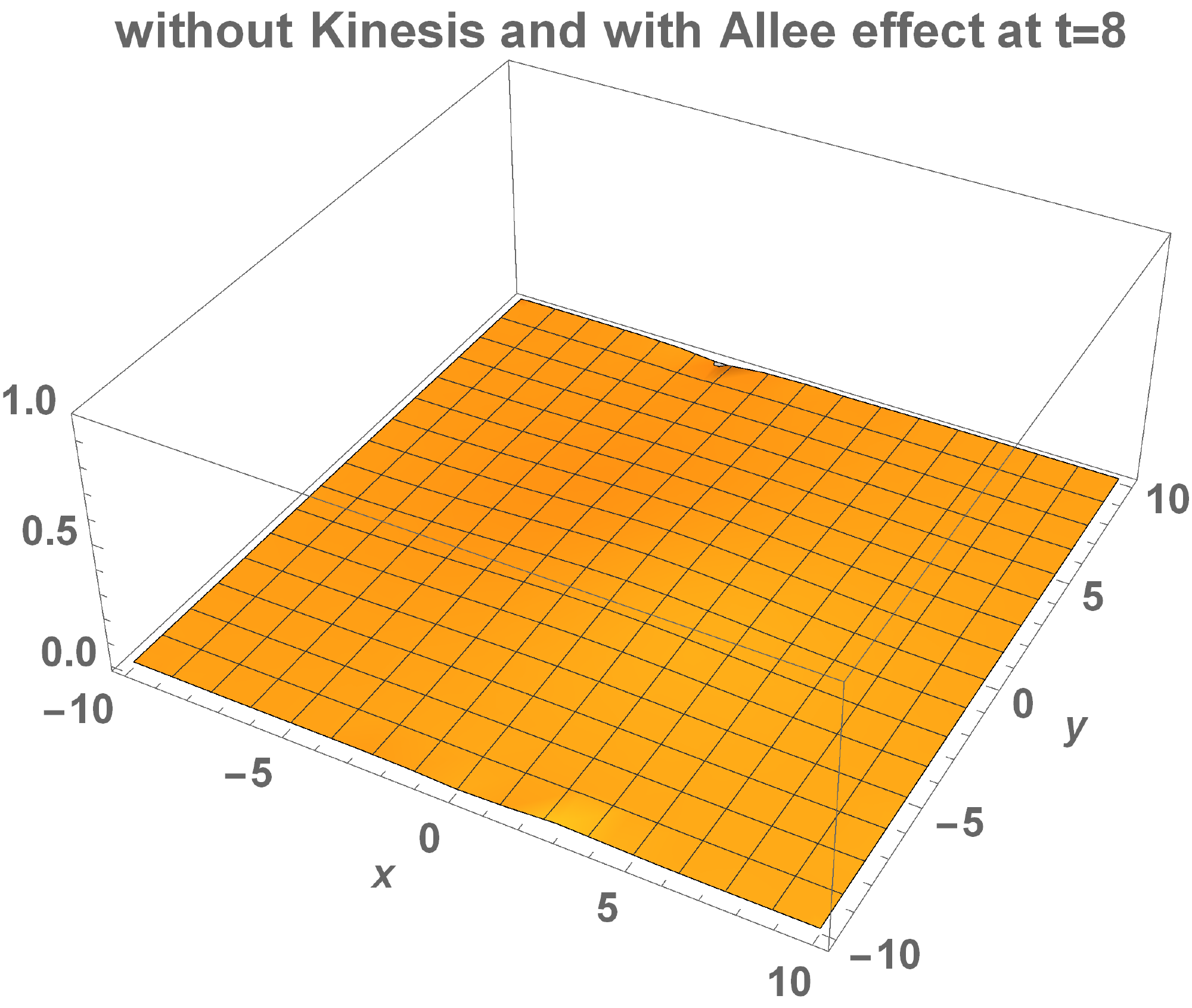}
c)\includegraphics[width=0.42\textwidth]{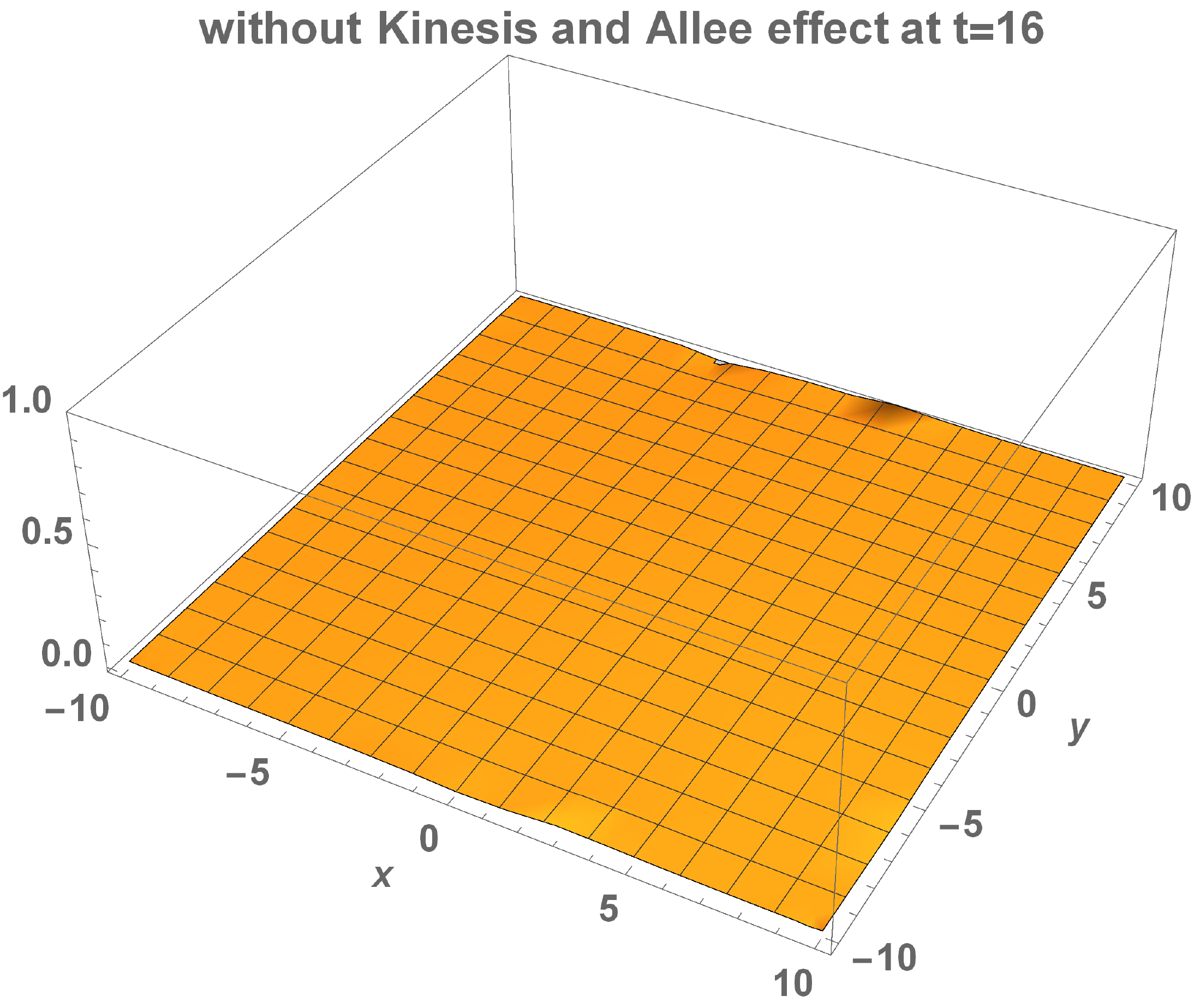}
\caption{ 2D Allee effect {\em without kinesis}  $\alpha =1$, $D=1$, $\beta=0.05$.\label{Alleewithoutkin}}
\end{figure}

\begin{figure}
\centering
a)\includegraphics[width=0.42\textwidth]{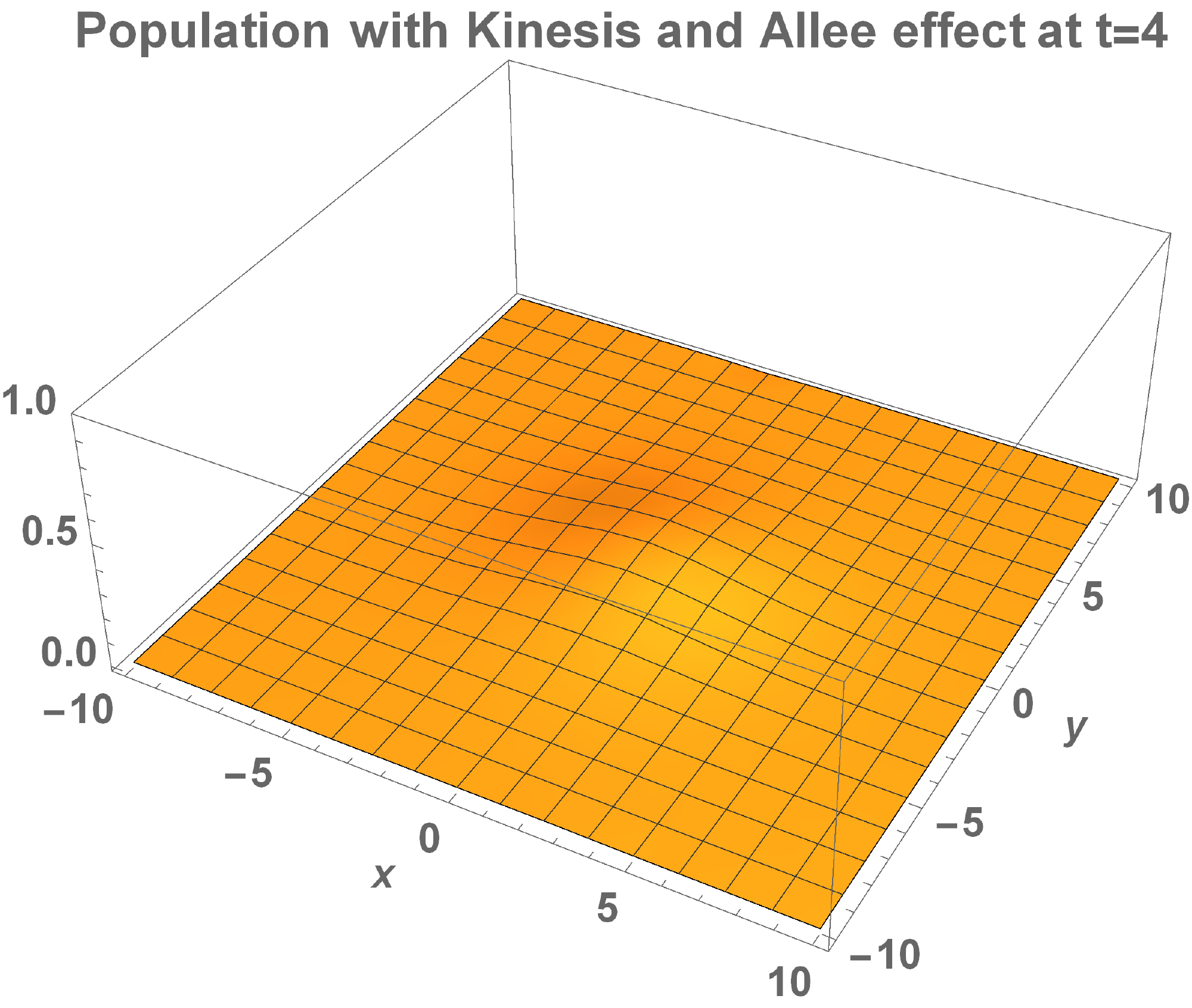}
b)\includegraphics[width=0.42\textwidth]{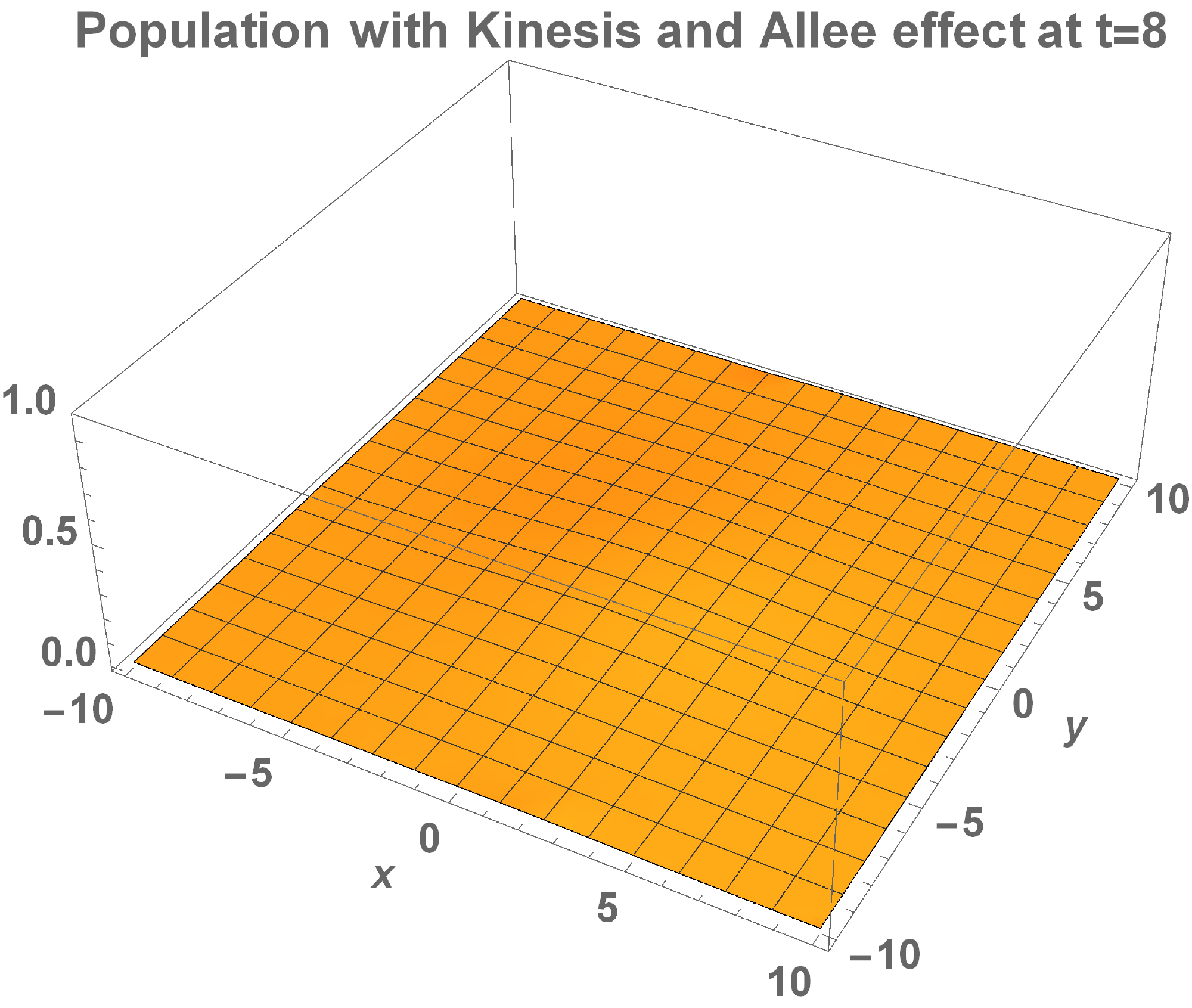}
c)\includegraphics[width=0.42\textwidth]{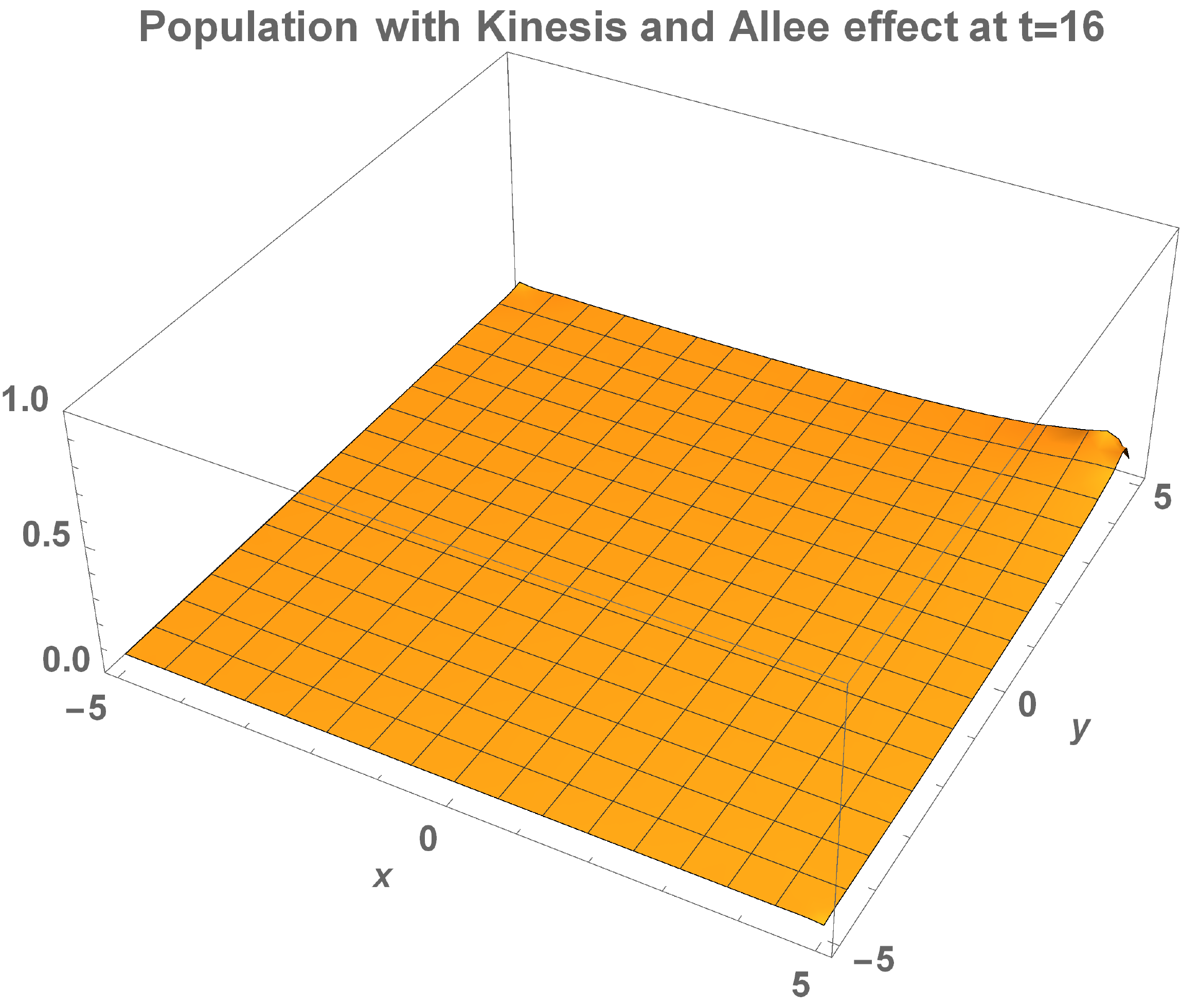}
\caption{ 2D Allee effect {\em with kinesis} $\alpha =1$, $D=1$, $\beta=0.05$.\label{Alleewithkin}}
\end{figure}

\begin{figure}
\centering
a)\includegraphics[width=0.4\textwidth]{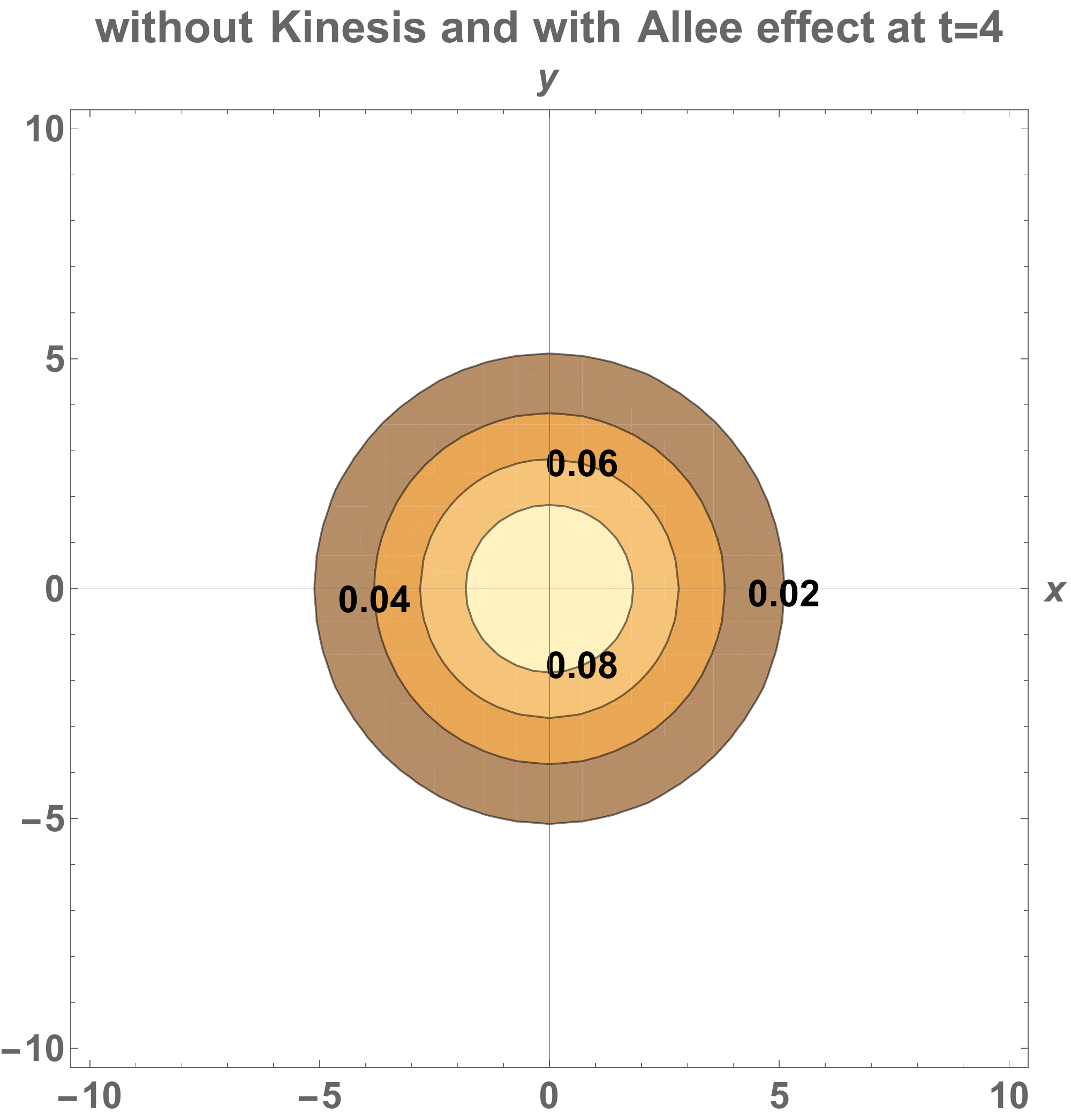}
b)\includegraphics[width=0.4\textwidth]{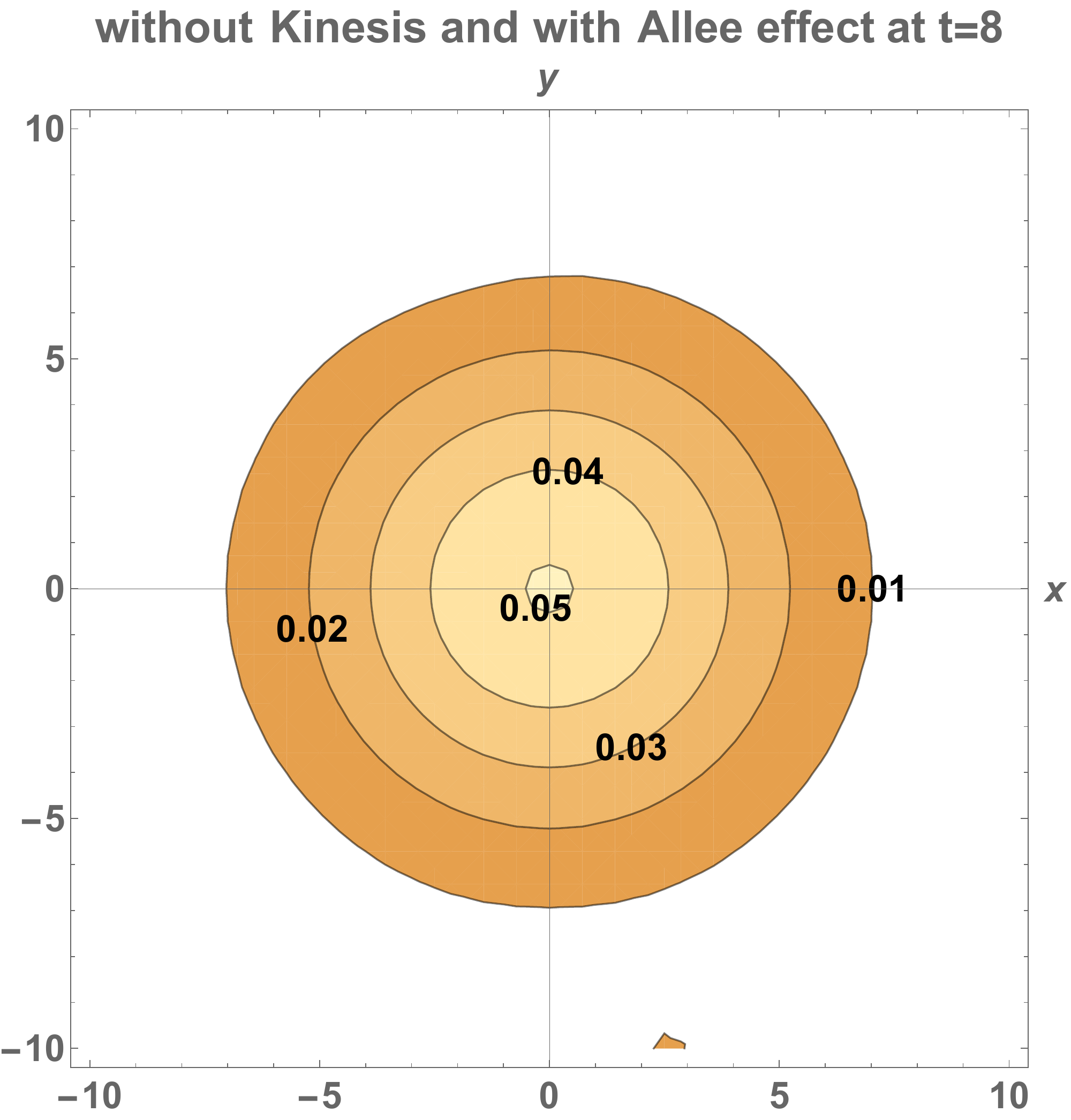}
c)\includegraphics[width=0.4\textwidth]{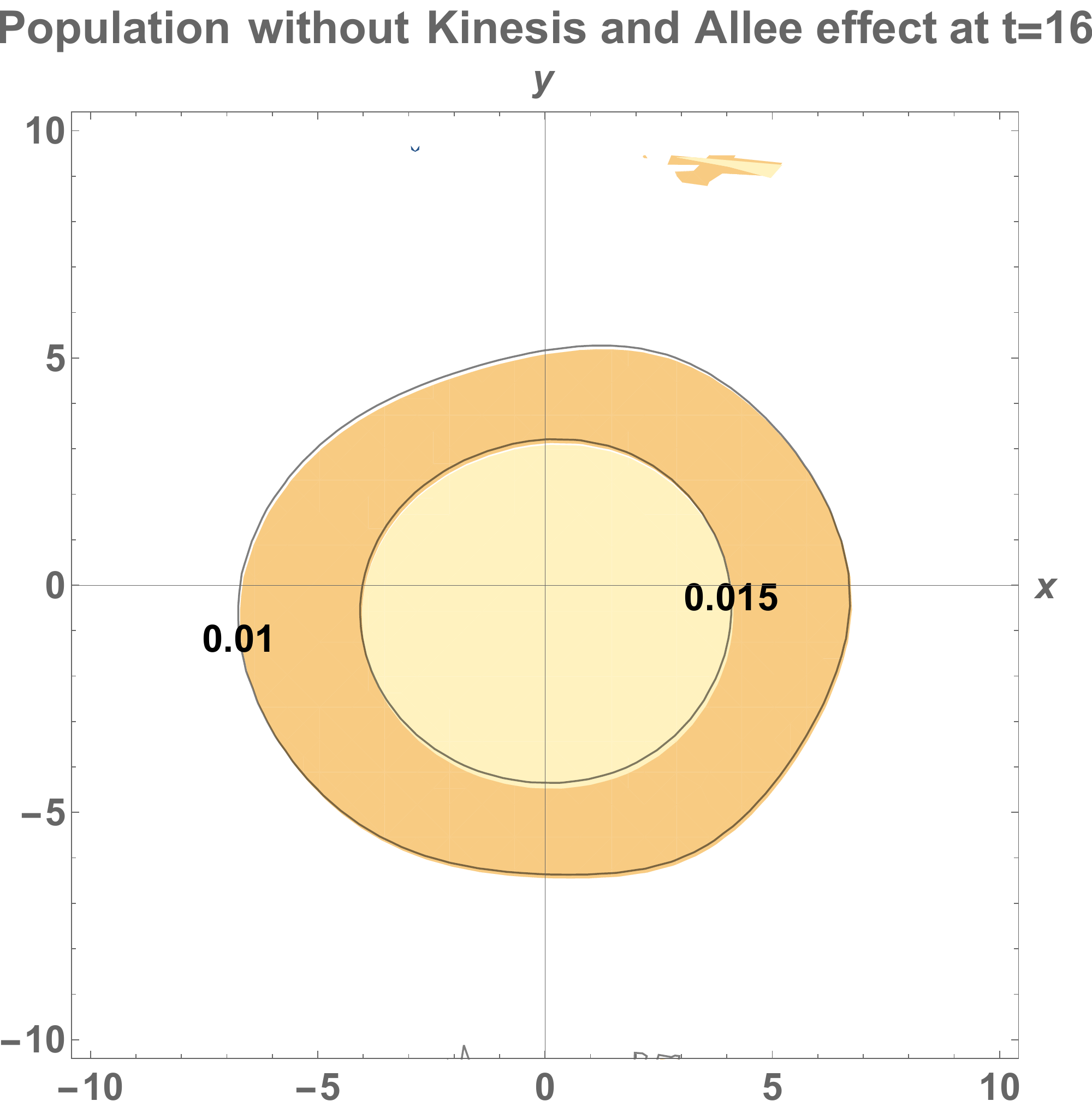}
\caption{ Invaded areas at center by the population {\em without kinesis} and Allee effect. The parameters are: $\alpha =1$, $D=1$, $\beta=0.05$.\label{CoAlleewithoutkin}}
\end{figure}

\begin{figure}
\centering
a)\includegraphics[width=0.42\textwidth]{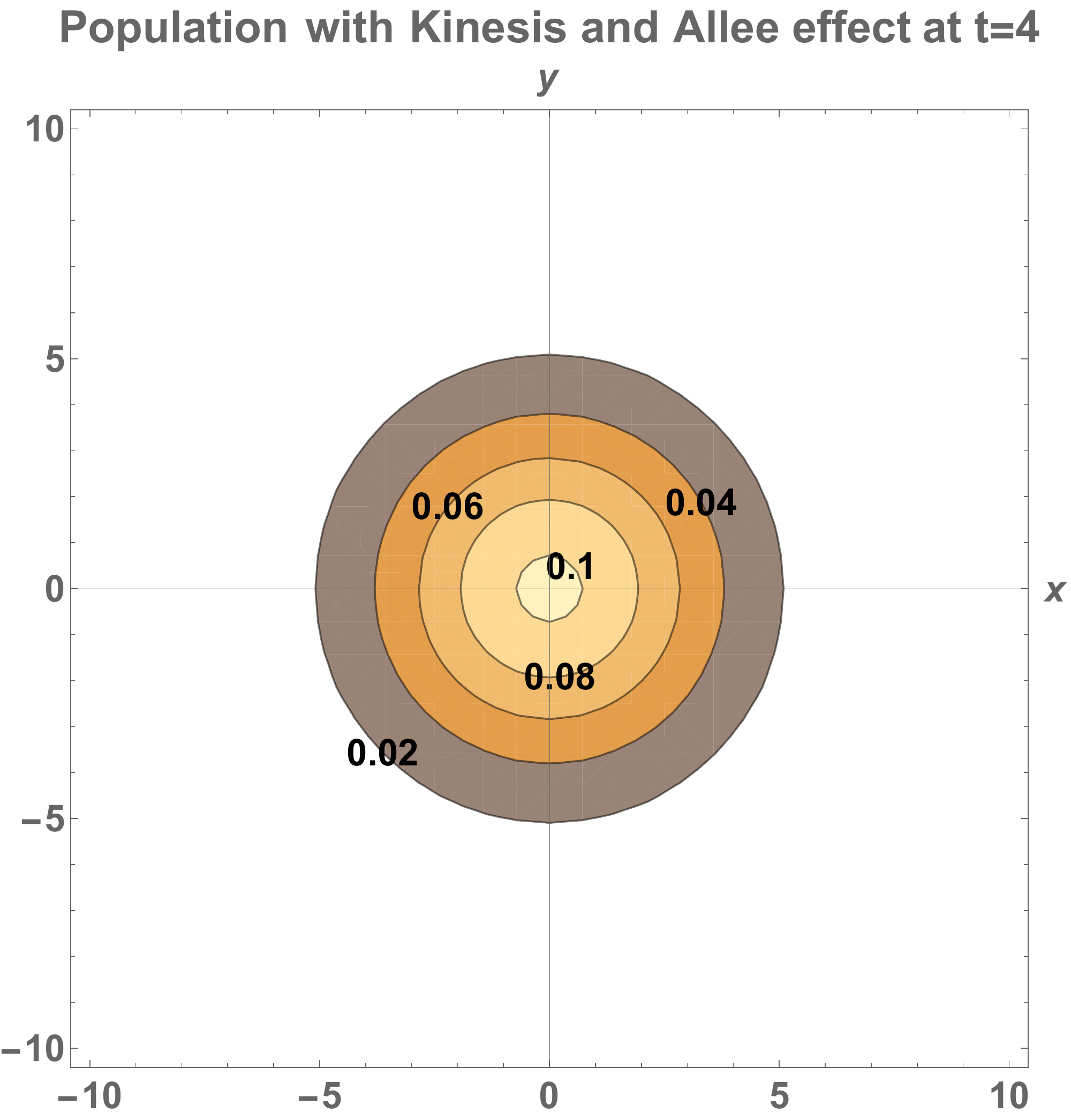}
b)\includegraphics[width=0.42\textwidth]{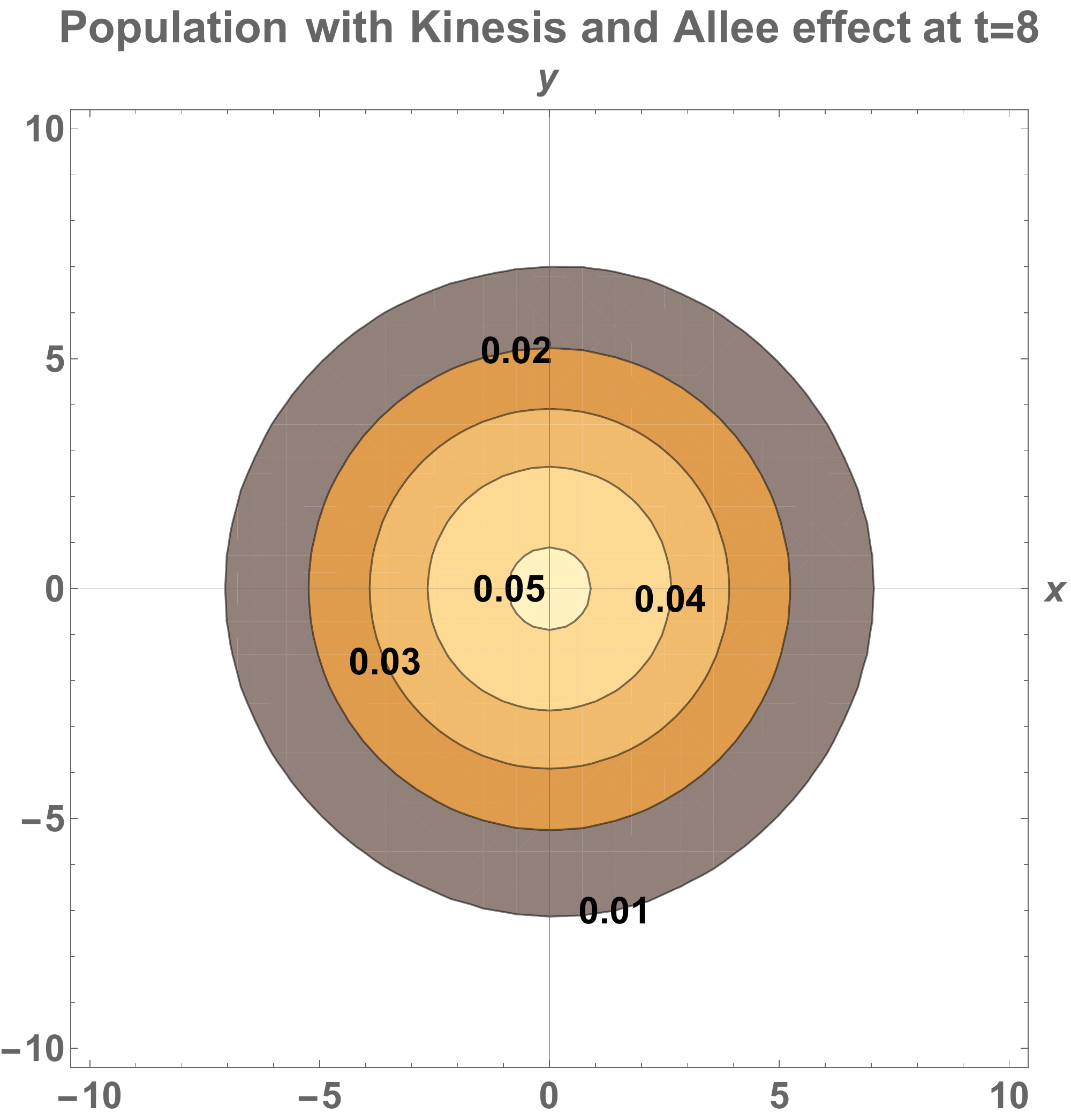}
\caption{  Invaded areas at center by the population {\em with kinesis} and Allee effect. At time $t=16$, the population dies out and there is no waves. The parameters are: $\alpha =1$, $D=1$, $\beta=0.05$.\label{CoAlleewithkin}}
\end{figure}

\begin{figure}
\centering
\subfigure[$t=4$]{\includegraphics[width=0.42\textwidth]{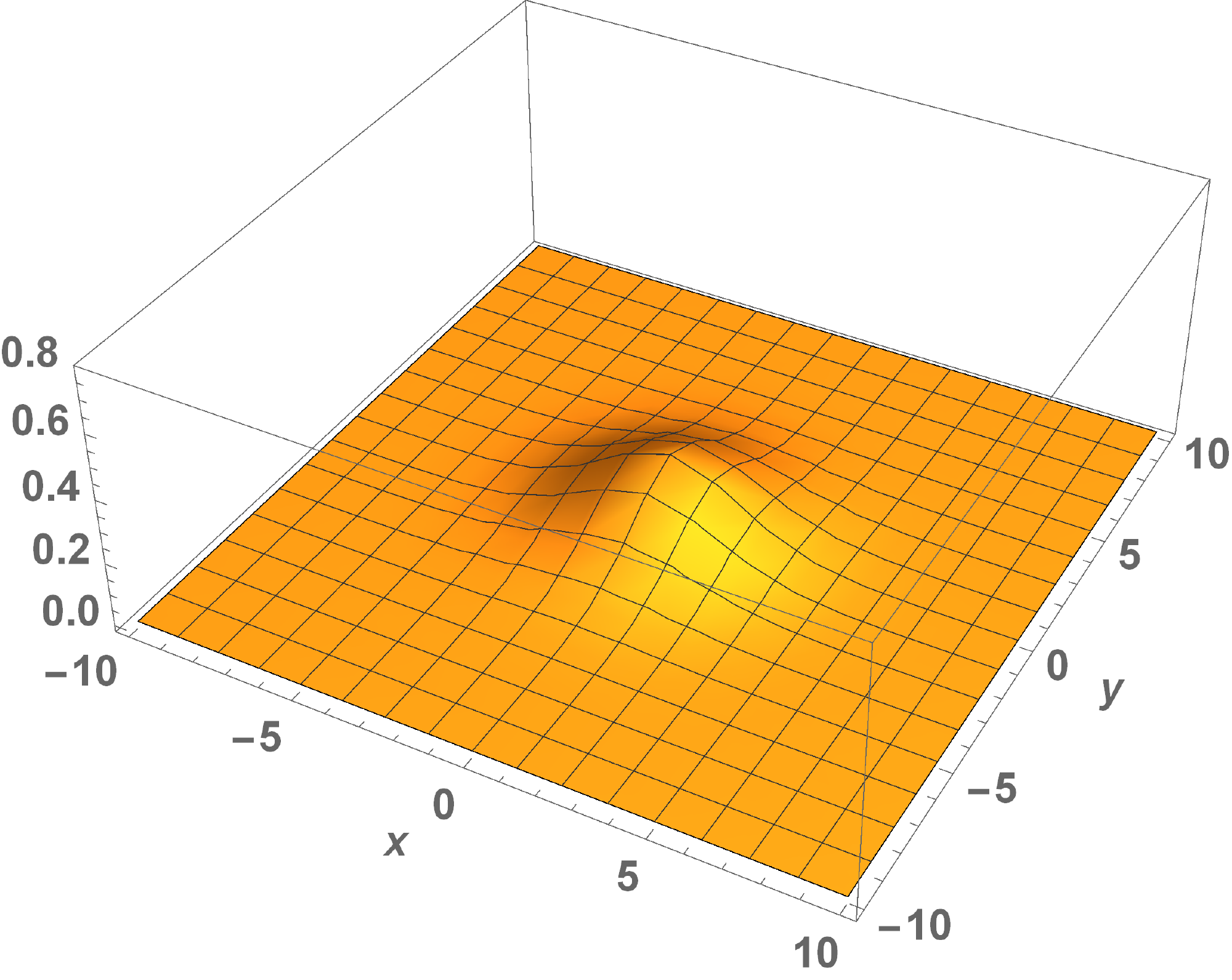}}
\subfigure[$t=8$]{\includegraphics[width=0.42\textwidth]{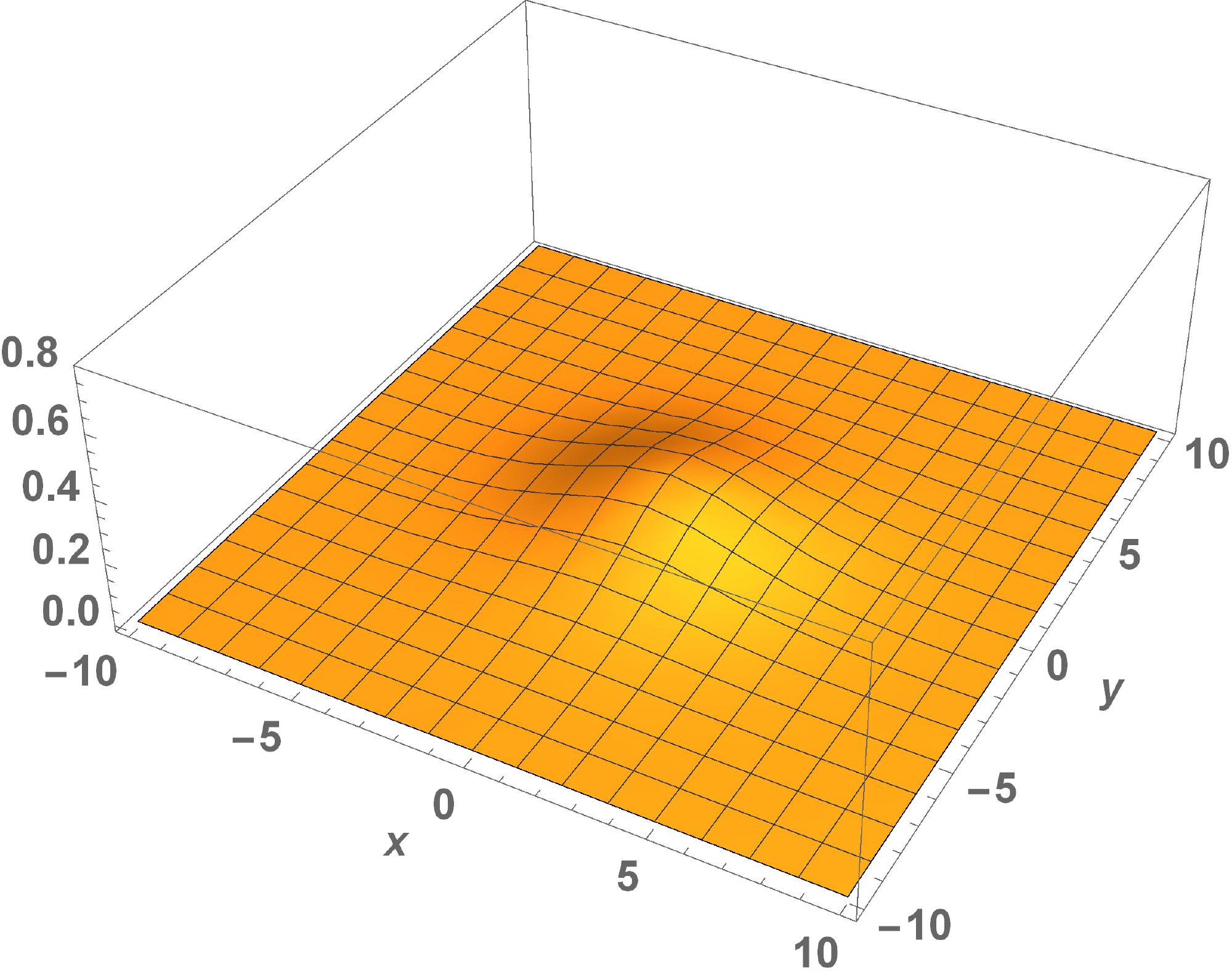}}
\subfigure[$t=16$]{\includegraphics[width=0.42\textwidth]{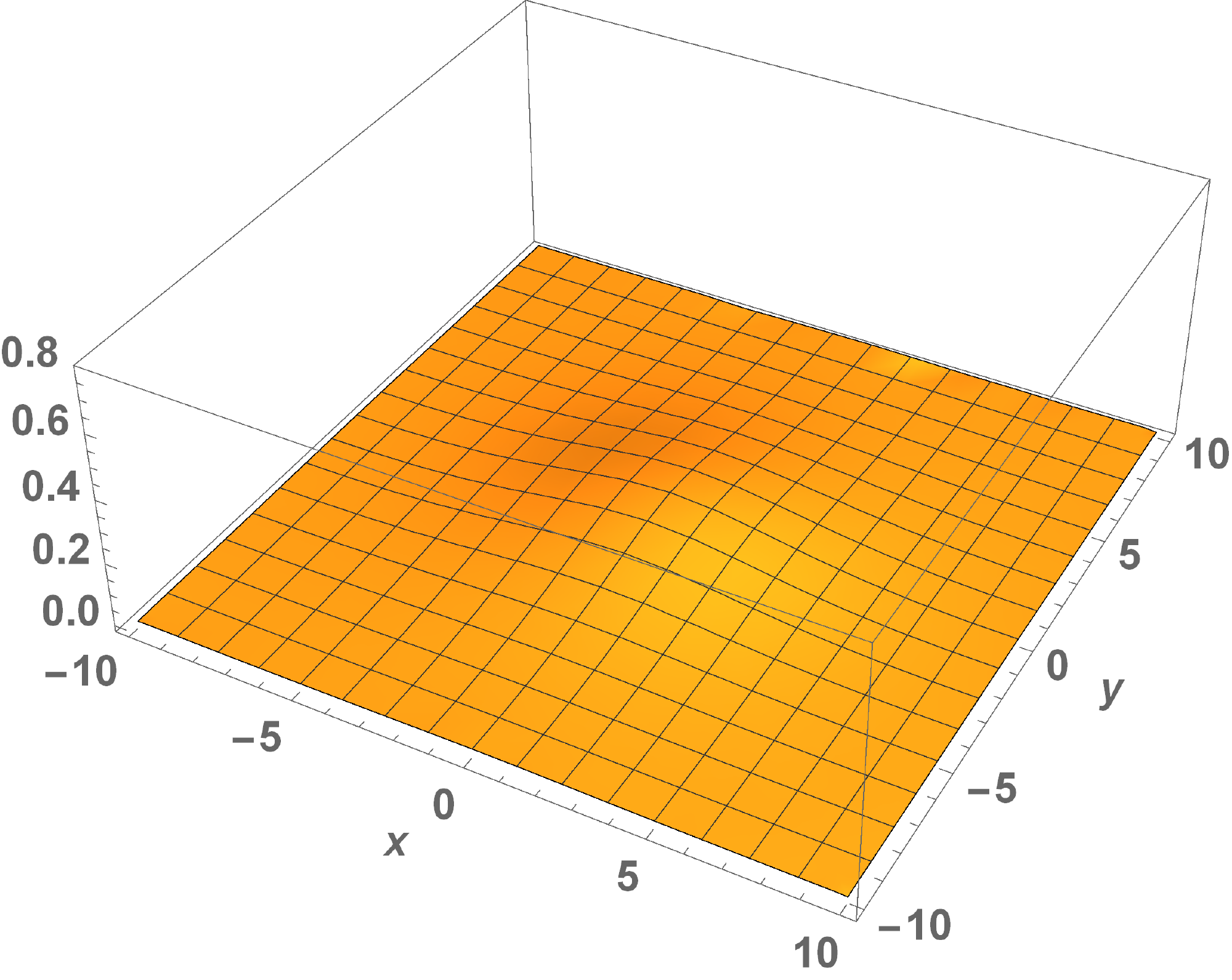}}
\caption{ 2D Allee effect {\em without kinesis} $\alpha =5$, $D=0.5$, $\beta=0.05$ with Neumann boundary conditions.\label{Alleewithoutkin2D}}
\end{figure}

\begin{figure}
\centering
\subfigure[$t=4$]{\includegraphics[width=0.42\textwidth]{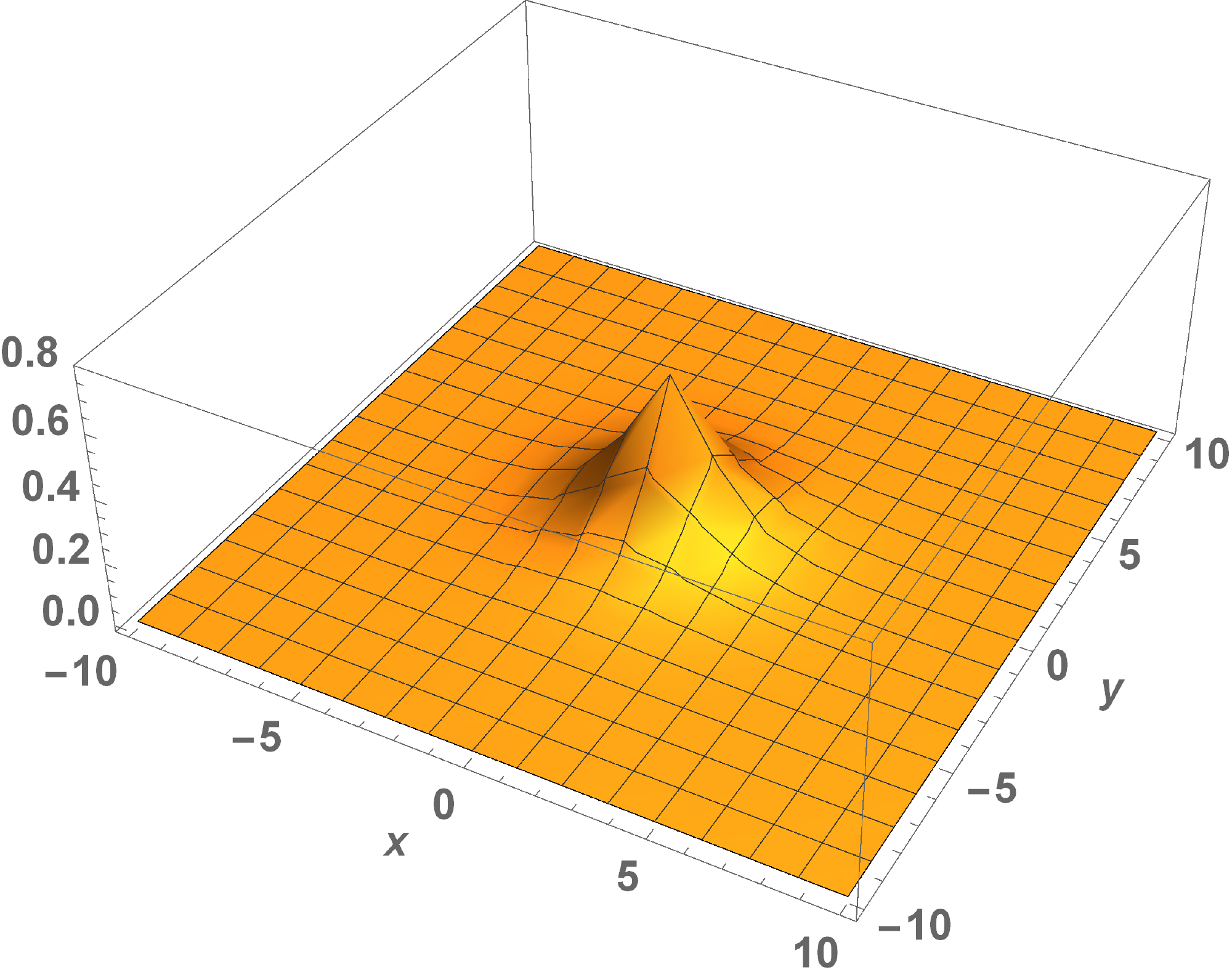}}
\subfigure[$t=8$]{\includegraphics[width=0.42\textwidth]{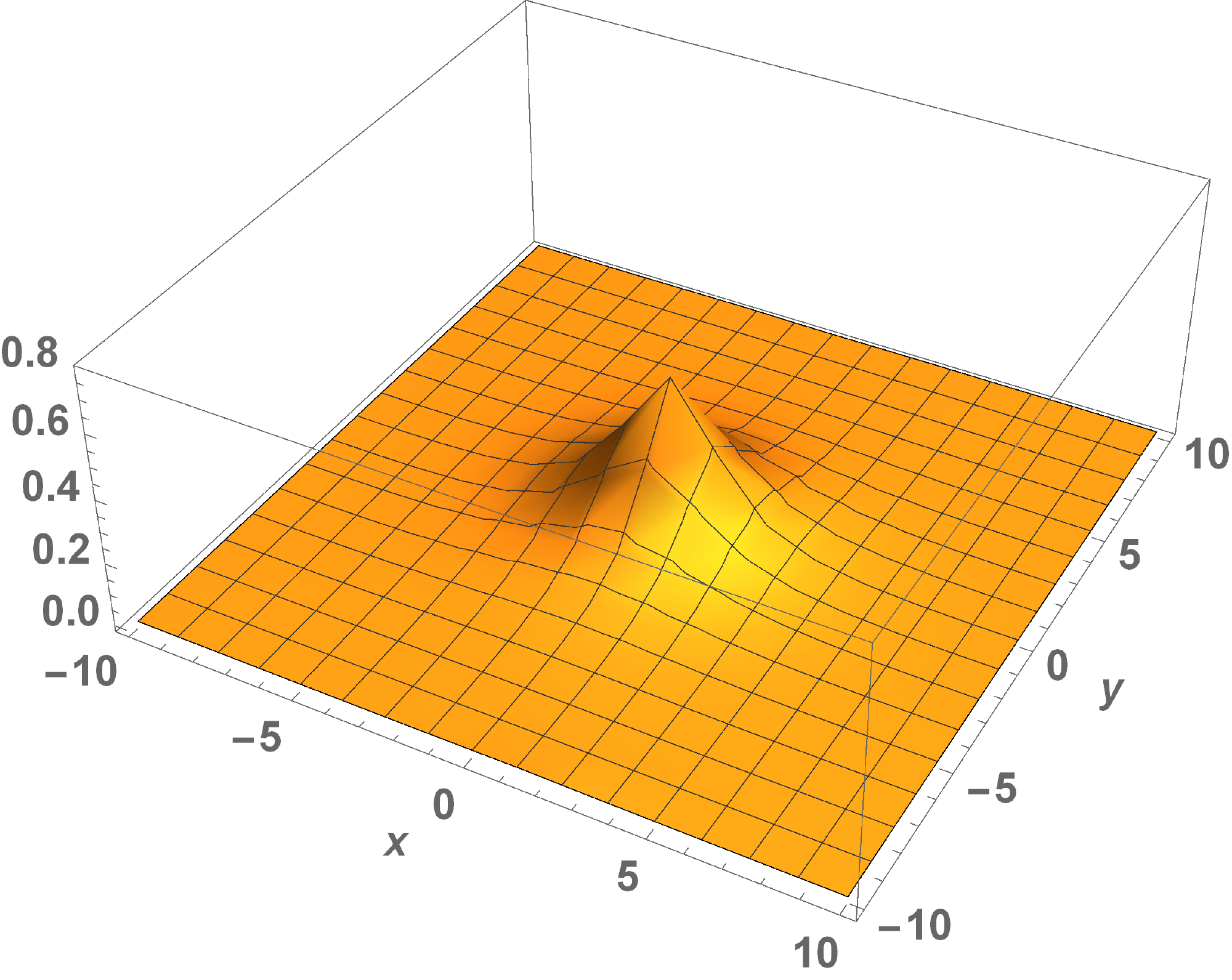}}
\subfigure[$t=16$]{\includegraphics[width=0.42\textwidth]{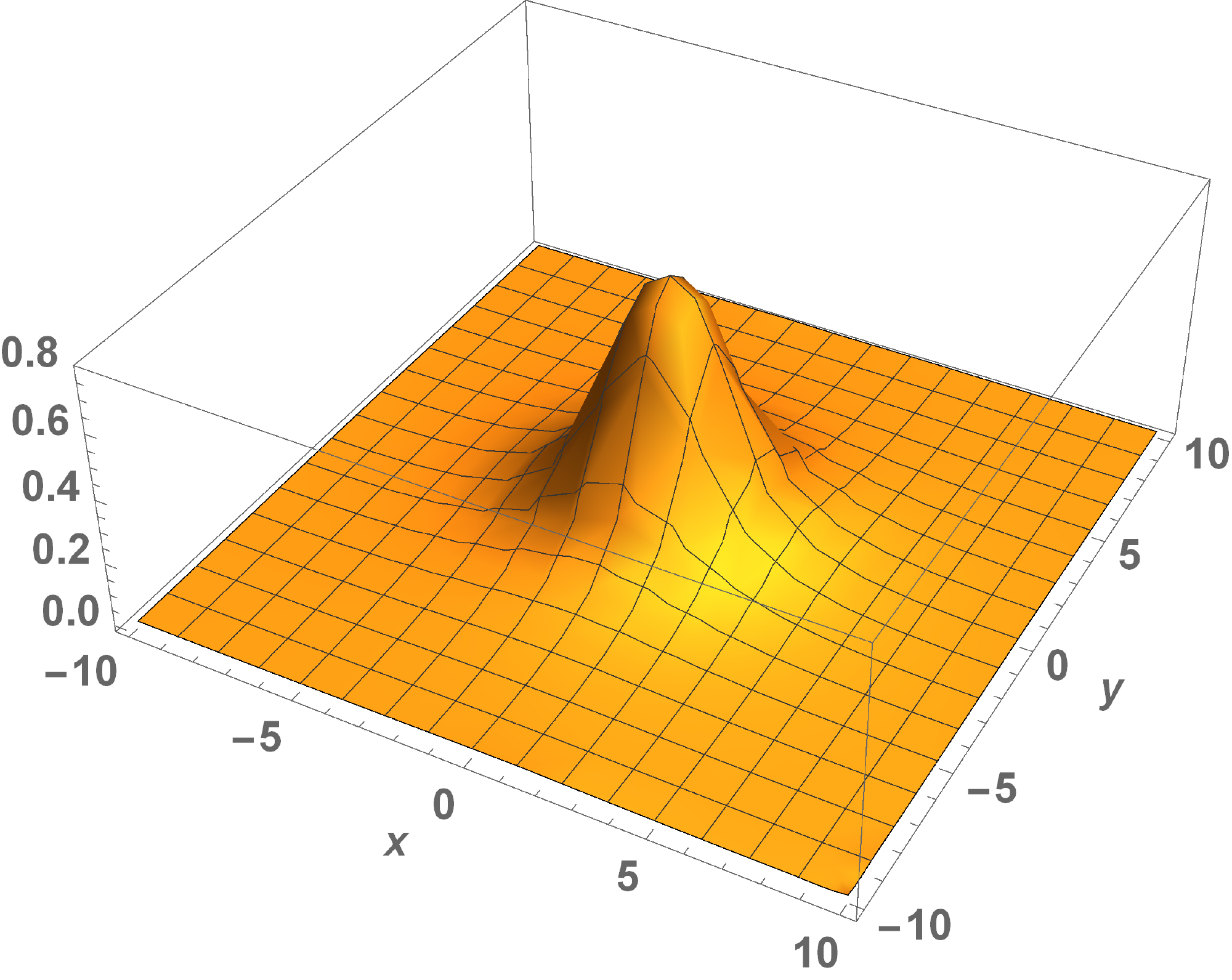}}
\caption{ 2D Allee effect {\em with kinesis} $\alpha =5$, $D=0.5$, $\beta=0.05$ with Neumann boundary conditions.\label{Alleewithkin2D}}
\end{figure}

\begin{figure}
\centering
\subfigure[$t=4$]{\includegraphics[width=0.42\textwidth]{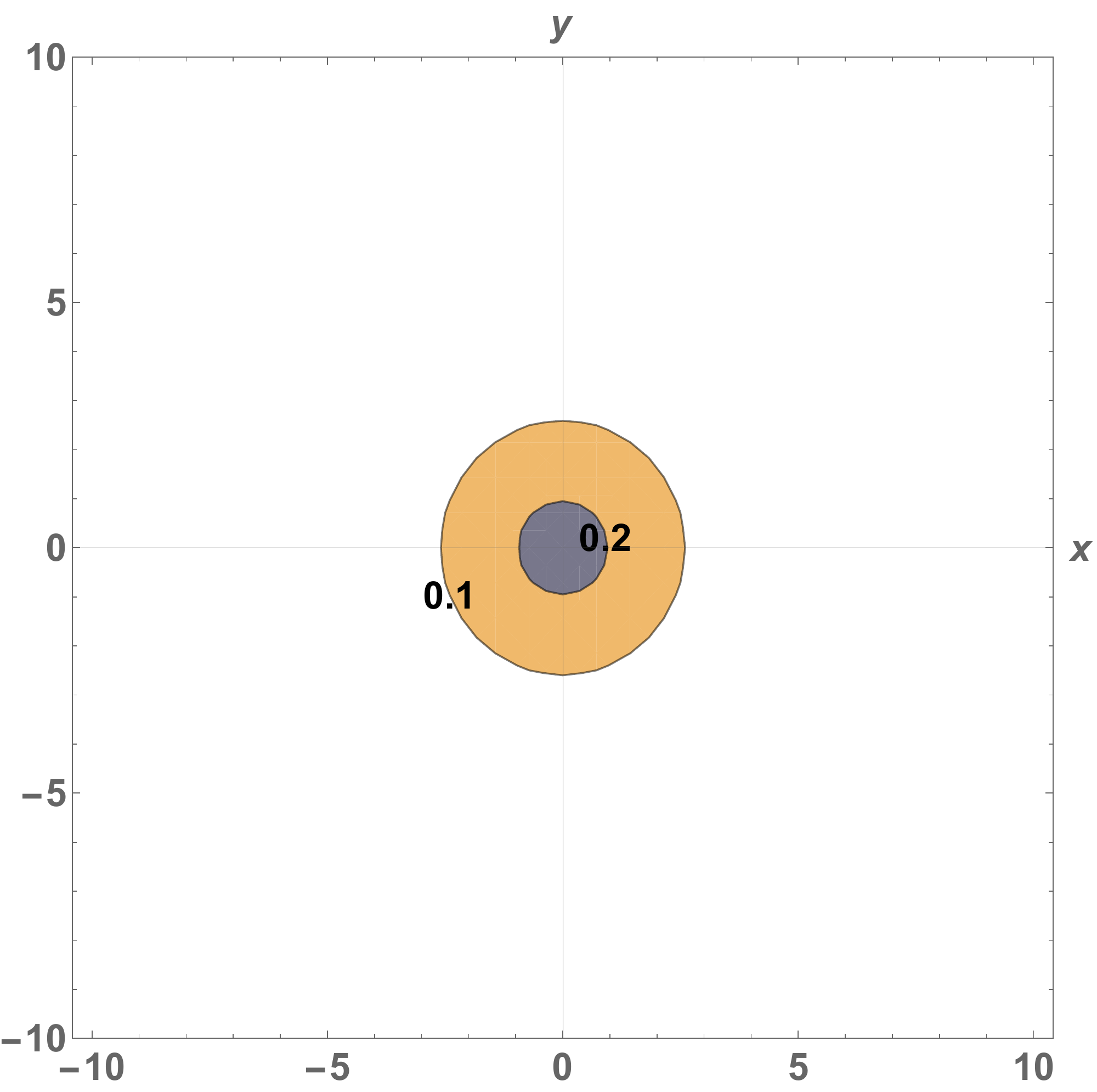}}
\subfigure[$t=8$]{\includegraphics[width=0.42\textwidth]{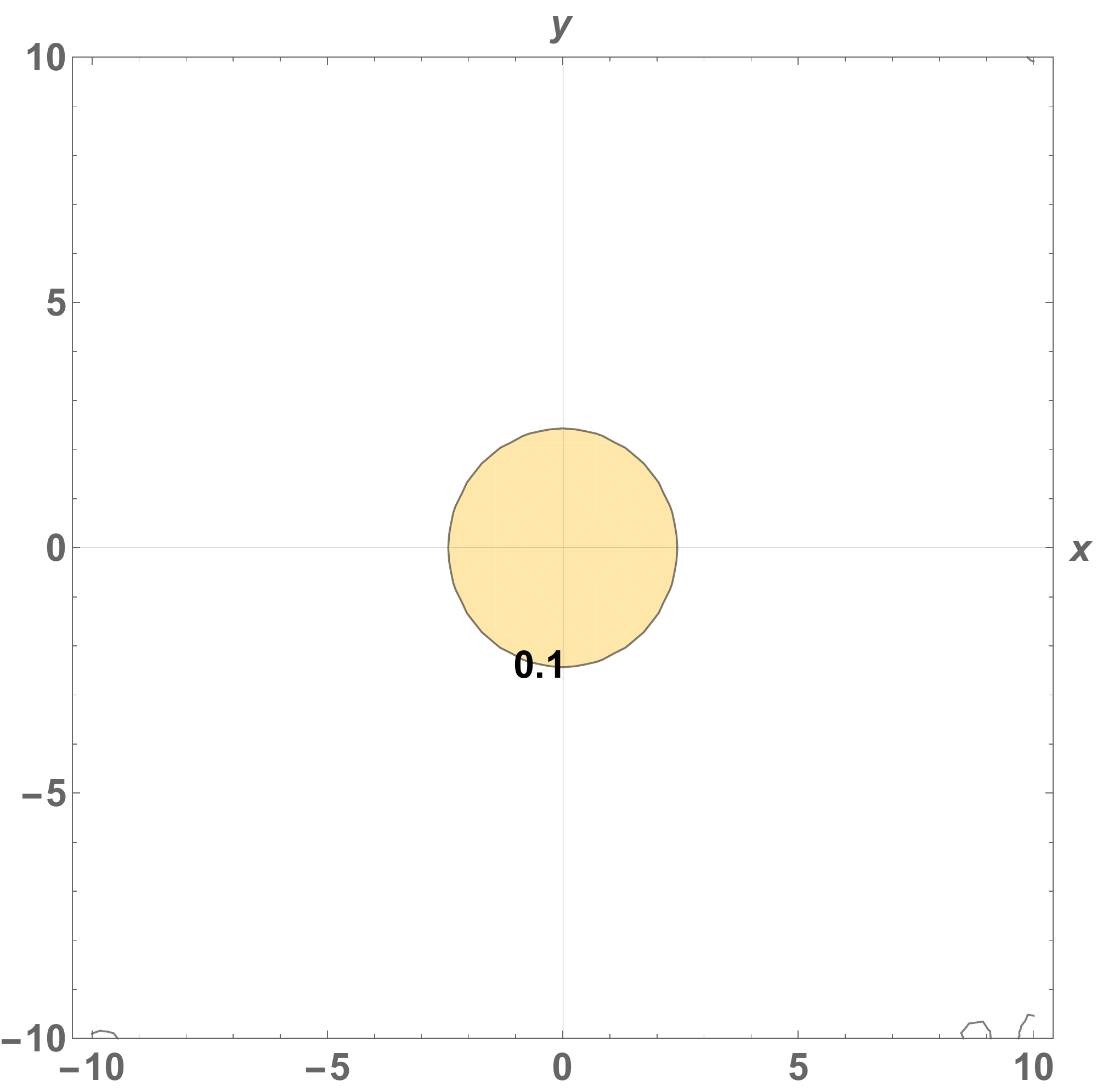}}
\caption{ Invaded areas at center by the population {\em without kinesis} and Allee effect. At time $t=16$, there will be no waves since the population dies. The parameters are: $\alpha=5$, $D=0.5$, $\beta=0.05$.\label{Copopwithoutkinesis}}
\end{figure}

\begin{figure}
\centering
\subfigure[$t=4$]{\includegraphics[width=0.38\textwidth]{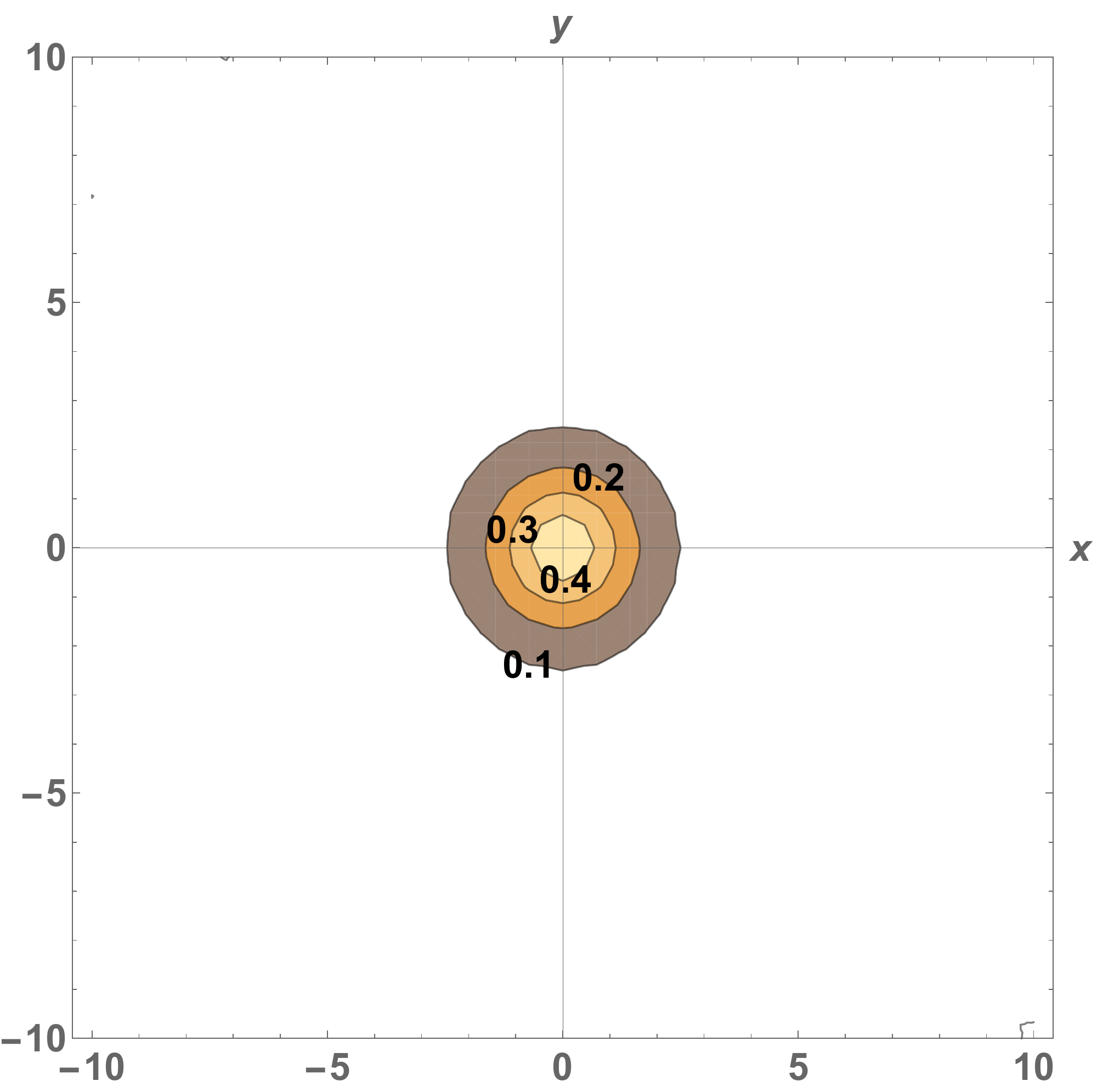}}
\subfigure[$t=8$]{\includegraphics[width=0.38\textwidth]{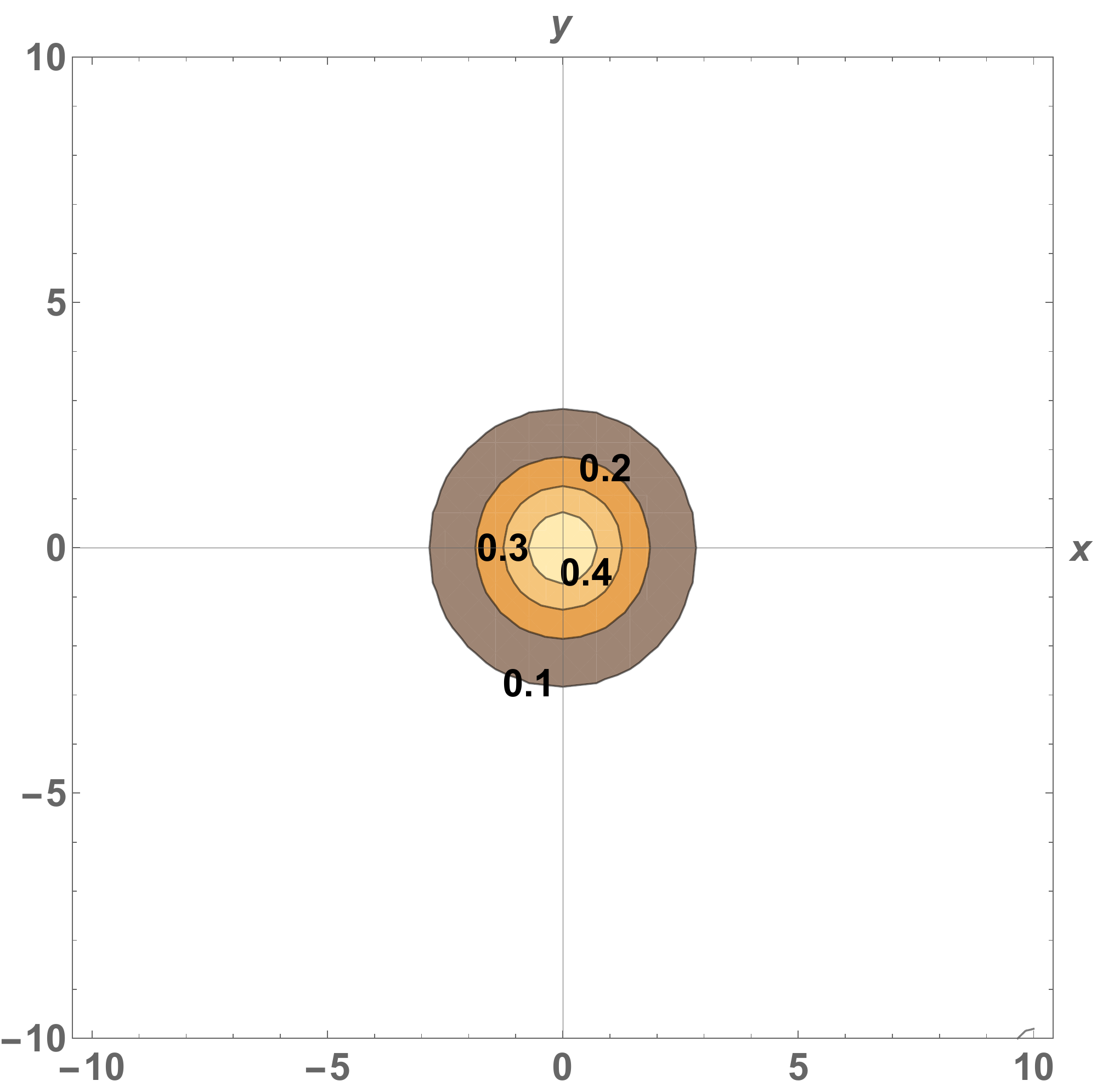}}
\subfigure[$t=16$]{\includegraphics[width=0.38\textwidth]{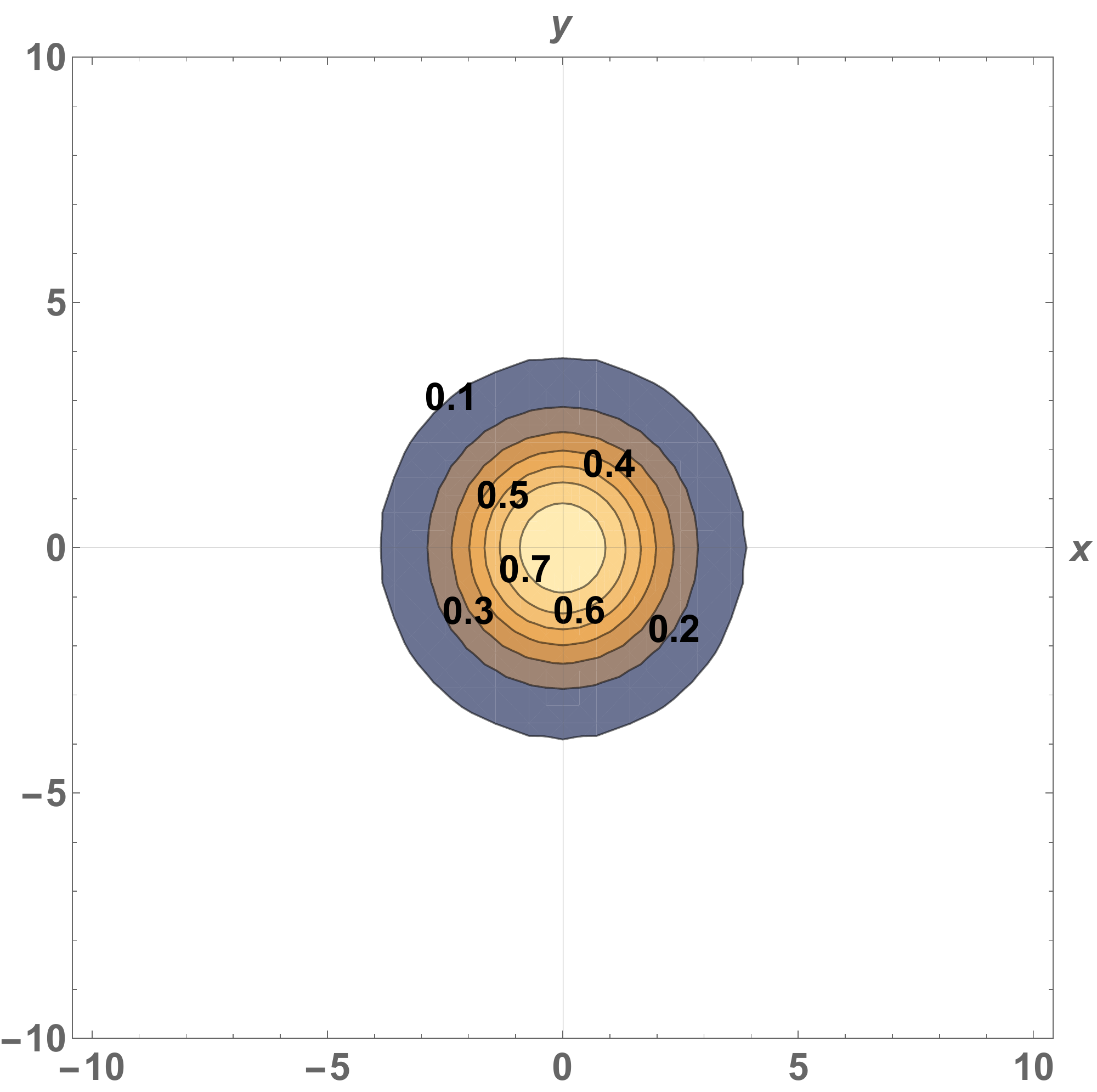}}
\caption{ Invaded areas at center by the population {\em with kinesis} and Allee effect. The parameters are: $\alpha =5$, $D=0.5$, $\beta=0.05$.\label{Copopwithkinesis}}
\end{figure}

\section{Predator-Prey Running Waves}

In this section, we consider competing species in the same environment. These two species interact with the model without kinesis and the new model with kinesis. Volpert and Petrovskii analysed the running waves of predator-prey interaction model in their study \citep{VolPet2009}. Now, we compare the population with kinesis and without kinesis as running wave behaviour.
\begin{itemize}
\item{ The PDE model for population with the constant diffusion coefficient (i.e. without kinesis):
\begin{eqnarray}
\partial_t u ( t,x)& = &D \nabla \cdot ( \nabla u ) + r_u  u( t,x),\label{predpreywithout1}\\
\partial_t v ( t,x)& = &D \nabla \cdot ( \nabla u ) + r_v v( t,x), \label{predpreywithout}
\end{eqnarray}
}

\item{  The PDE model for population  with kinesis:

\begin{eqnarray}
\partial_t u ( t,x)& = &D \nabla \cdot \left(e^{-\sigma_u r_u   } \nabla u \right) + r_u  u( t,x), \label{predpreywith1}\\
\partial_t v ( t,x)& = &D \nabla \cdot \left( e^{-\sigma_v  r_v} \nabla u \right)  +  r_v v( t,x). \label{predpreywith}
\end{eqnarray}
}
\end{itemize}
We consider the simplest predator-prey model in the classical Lotka-Volterra form in dimensionless variables $r_u= k-v$ and $r_v=u-m$; $u$ and $v$ are the population densities of prey and predator.

Eqs. (\ref{predpreywithout1}), (\ref{predpreywithout}),  (\ref{predpreywith1}) and   (\ref{predpreywith}) are solved for one space variable $x\in[-100,100]$  with  Dirichlet boundary  conditions and with the initial conditions:
\begin{equation}
u( 0,x) = v( 0,x) =\dfrac{1}{1+e^{\lambda x}}; \lambda=10.
\end{equation}

\begin{figure}[!ht]
\centering
a)\includegraphics[width=0.42\textwidth]{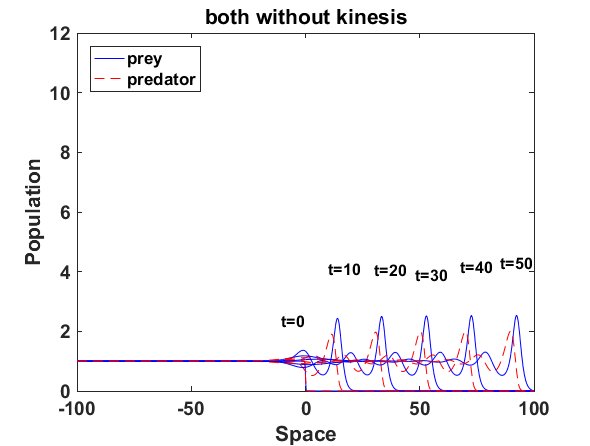}
b)\includegraphics[width=0.42\textwidth]{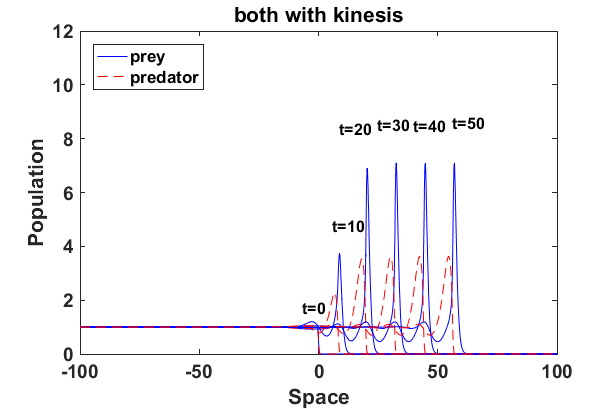}
\caption{Travelling waves of predator-prey diffusion (a) for animals without kinesis,(\ref{predpreywithout1}, \ref{predpreywithout}) $\sigma_1 = \sigma_2 = 1$ (b) for animals with kinesis (\ref{predpreywith1}, \ref{predpreywith}) $\sigma_1 = \sigma_2 = 1$. The values of constants are: $D=1$, $k=1$, $m=1$. \label{predpreytravelling1}}
\end{figure}

\begin{figure}[!ht]
\centering
a)\includegraphics[width=0.42\textwidth]{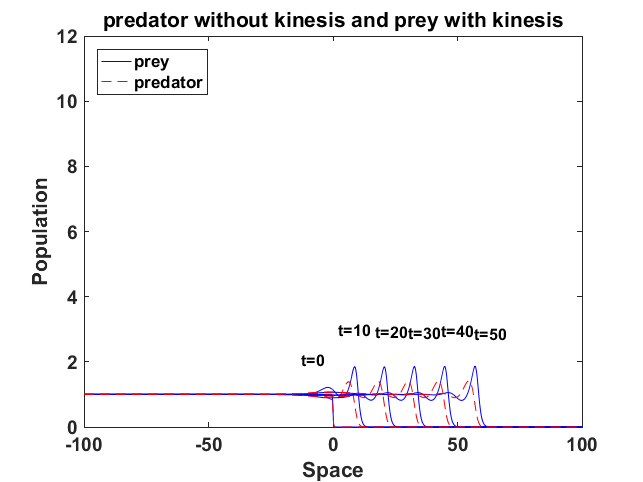}
b)\includegraphics[width=0.42\textwidth]{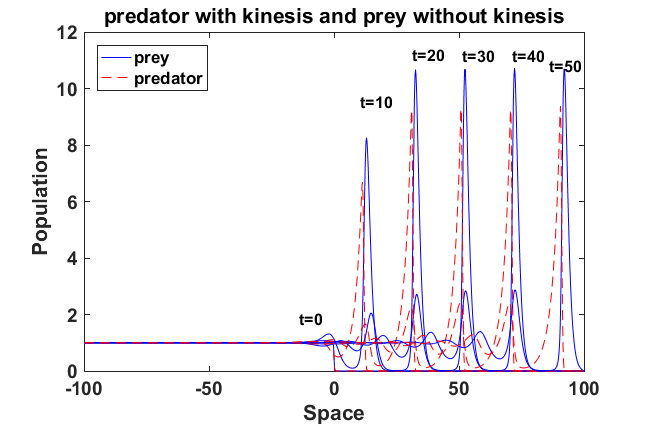}
\caption{Travelling waves of predator-prey diffusion (a) for predators with kinesis and preys without kinesis $\sigma_1 =0,$ $\sigma_2 = 1.$ (b) for predators without kinesis and preys with kinesis $\sigma_1 =1,$ $\sigma_2 = 0.$ The values of constants are: $D=1$, $k=1$, $m=1$. \label{predpreytravelling2}}
\end{figure}

The numerical simulations with the 1D case show that depending on the values the systems (\ref{predpreywithout1}), (\ref{predpreywithout}), (\ref{predpreywith1}) and   (\ref{predpreywith}) show a variety of travelling fronts.

We compare four cases with Figs.~\ref{predpreytravelling1} and \ref{predpreytravelling2}:
\begin{itemize}
\item both predator and prey without kinesis,
\item both predator and prey with kinesis,
\item predator without kinesis and prey with kinesis,
\item predator with kinesis and prey without kinesis.
\end{itemize}

All of them have the running wave behaviour. We demonstrate with Fig.~\ref{predpreytravelling1}b that the predator and prey populations with kinesis have the running wave behaviour and the peaks between predator and prey increase. For the predator without kinesis and prey with kinesis (Fig.~\ref{predpreytravelling2}a), the amplitude decreases. On the other hand, the predator with kinesis and prey without kinesis (Fig.~\ref{predpreytravelling2}b) running wave peaks of the population density become sharper.


\section{Discussion}

It has been suggested a purposeful kinesis model that the diffusion coefficient depends on well-being which is measured by the reproduction coefficient by \cite{GorCabuk2018}. The system formalizes the simple rule: ``Let well enough alone". Volpert and Petrovskii presented the study on travelling waves of the reaction-diffusion model \cite{VolPet2009}.

In this paper, we analysed the running wave behaviour of the reaction-diffusion model with kinesis. The running wave theory begins with \cite{Fisher1930} and Kolmogorov, Petrovskii and Piskunov \citep{KPP1937}. They defined the travelling wave solutions of the scalar reaction-diffusion equation and analysed their existence, stability and the speed of these waves. Lewis and Kareiva explored the relevance between the Allee effect parameter $\beta$ and the relative wave speed. They also studied on the two-dimensional spread of invading populations with Allee effect.

In this study, we have demonstrated the speed of the running waves of the population with kinesis. These following have been observed:

\begin{itemize}

\item While the relation between the diffusion coefficient and reproduction coefficient $\alpha$ is increasing the running wave speed monotonically decreases, and at some point stabilizes (see Fig.~\ref{velocityalpha}).

\item The Allee effect parameter $\beta$ and velocity $v$ have been displayed by Fig.~\ref{speedbeta} for the values of normalization coefficient $k$ on the running wave front $u=0.5$. While the Allee effect parameter is increasing, the populations' running speed decreases. When $\beta = 0.5$, the velocity is 0. After that value of $\beta$, the waves of the population model show the running wave behaviour through the left hand side. The individuals with kinesis run slower than the population without kinesis. This may lead up an extinction in the invaded area.

\item The initial distributions of the invaded area are the same for both populations (see Fig.~\ref{initial}). System stabilization is observed by $r=0$. Therefore, Kinesis does not change the stability of homogeneous positive steady-states.

\item The invasion of the population with kinesis occupies the larger area than without kinesis. That is, they stay alive in the habitat.

\item Allee dynamics can affect the population extinction (see Fig.~\ref{Copopwithoutkinesis}). On the other hand, with the higher diffusion, the population with kinesis do not extinct and can invade a large area (see Fig.~\ref{Copopwithkinesis}).

\item We have also presented the running wave behaviour of the predator-prey model with kinesis. The constant diffusion model waves were recently analysed by \cite{VolPet2009} for the predator-prey model. We have observed also that this model can have running wave behaviour. Kinesis of prey decreases amplitude of the wave smaller, whereas kinesis of the predators can significantly increase the peak value (Fig.~\ref{predpreytravelling2}b).

\end{itemize}

These population models with kinesis can be adapted to several models. It can be seen that individuals may have the running wave behaviour in the invaded areas.

\section*{Acknowledgement} I am grateful to A.N. Gorban for his advises.

\end{document}